\documentclass[12pt]{article}
\usepackage{latexsym,amssymb,amstext,amsmath,array}
\usepackage{amscd,bbm}
\usepackage{cite}
\usepackage{slashed}
\usepackage{mathrsfs}
\usepackage{hyperref}
\usepackage{xcolor}

\newcommand {\cA}{{\cal A}}

\newcommand {\cJ}{{\cal J}}

\newcommand {\cN}{{\cal N}}

\newcommand {\cR}{{\cal R}}
\newcommand {\cS}{{\cal S}}

\newcommand{\bR}{{\bf R}}

\def\a{\alpha}
\def\b{\beta}

\def\G{\Gamma}

\def\l{\lambda}

\def\o{\omega}

\def\z{\zeta}

\def\U{\Upsilon}

\newcommand{\be}{\begin{equation}}
\newcommand{\ee}{\end{equation}}
\newcommand{\bea}{\begin{eqnarray}}
\newcommand{\eea}{\end{eqnarray}}

\newcommand{\ba}{\begin{array}}
\newcommand{\ea}{\end{array}}

\def\double #1{#1{\hbox{\kern-2pt $#1$}}}

\newcommand{\bsubeq}{\begin{subequations}}
\newcommand{\esubeq}{\end{subequations}}

\begin{document}

\begin{titlepage}

\begin{center}

\vskip .3in \noindent

{\Large \bf{Maximally \emph{$\cN$}-extended super-BMS$_3$ algebras\\[1mm] and
    Generalized 3D Gravity Solutions}}

\bigskip

{Nabamita Banerjee}$^{\,a,}$\footnote{nabamita@iiserpune.ac.in (on lien from IISER Pune)}, {Arindam Bhattacharjee}$^{\,b,}$\footnote{arindam.bhattacharjee@students.iiserpune.ac.in}, {Ivano Lodato}$^{\,c,} $\footnote{ivano@ilodato@fudan.edu.cn}, \\ {Turmoli Neogi}$^{\,b,}$\footnote{turmoli.neogi@students.iiserpune.ac.in}\\

\bigskip
$^{a}$ \em Indian Institute of Science Education and Research Bhopal\\
Bhopal Bypass, Bhauri, Bhopal 462066 \\
$^{b}$ \em Indian Institute of Science Education and
Research Pune,\\ Homi Bhabha Road, Pashan, Pune 411 008, India \\
$^{c}$ \em Department of Physics and Center for Field Theory and Particle Physics,\\
Fudan University, 200433 Shanghai, China \\

\vskip .5in
{\bf Abstract }
\vskip .2in
\end{center}
We consider the maximal $\cal{N}-$extended supergravity theory in 3 dimensions with fermionic generators transforming under real but non necessarily irreducible representations of the internal algebra. We obtain the symmetry algebra at null infinity preserving boundary conditions of asymptotically flat solutions, i.e. the maximal $\cal{N}-$extended super-BMS$_3$ algebra, which possesses non-linear correction in the anti-commutators of supercharges. We present the supersymmetric energy bound and derive the explicit form of the asymptotic Killing spinors. We also find the most generic circular symmetric  ground state of the theory, which corresponds to a non-supersymmetric cosmological solutions and derive their entropy.  
.

\vfill
\eject

\end{titlepage}

\newpage
\tableofcontents
\section{Introduction and Summary}
(Super)gravity theories in three dimensions have many
interesting  features. First and foremost, the vanishing of the Weyl
tensor in 3D implies a decomposition of the Riemann tensor $R_{\mu \nu
\rho \sigma}$ into the Ricci tensor $R_{\mu \nu}$
and Ricci scalar $R$. Solutions with zero cosmological
constant hence are always locally (in a neighborhood of any point) Minkowski spacetime, since they satisfy the dynamical equation $R_{\mu \nu}=0$. Thus 
asymptotically flat solutions of Einstein's equations in 2+1 dimensions possess 
no local degrees of freedom. In other words, gravitational radiation (or propagating gravitons) are not solutions of the classical (or quantum) theory\footnote{ Same conclusions can be reached for solutions with non-zero cosmological constant, which are locally all isomorphic to (A)dS$_3$.}.\\
Nevertheless, a large variety of gravitational solutions exists whenever global topological structures, such as the holonomy of the manifold are considered.   
If the holonomy of the spacetime is trivial, then a single coordinate patch parametrizing the neighborhood of a point with metric $\eta_{\mu\nu}$ can be extended globally.
If the holonomy is non-trivial as non-contractible cycles exist in the manifold, a single coordinate patch fails to cover the whole spacetime. Thus in this case the global solution differs from $\eta_{\mu\nu}$. 
Hence solutions of 3 dimensional (super)gravity can be classified by their holonomy structure. A detailed analysis can be found in \cite{Carlip:1995zj} and references there in.\\
Other interesting features of asymptotically flat solutions of 3D (super)gravity are related to its asymptotic structure. Whenever asymptotically flat spacetimes with non-trivial holonomies (or equivalently conical singularities) are considered it is not possible to define linear momentum and supercharges at spacelike infinity \cite{PhysRevD.29.2766, 0264-9381-1-1-001}. On the other hand it is well known that one can impose a set of relaxed boundary conditions on any $\cN$ supersymmetric asymptotically AdS$_3$ solution such that the symmetry algebra at spacelike infinity is enhanced from SO$(2,2|2\cN)$ to the $(\cN,\cN)$ super-Virasoro algebra. Similar symmetry enhancements occur for 3D flat gravity at null infinity, where the group of asymptotic symmetries is enhanced to the infinite dimensional BMS$_3$ symmetries \cite{Bondi:1962px,Sachs:1962zza}, specifically the semi-direct product of supertranslations (translations along the null-time coordinates), super-rotations (conformal reparametrizations of the circle)  \cite{Ashtekar:1996cd, Barnich:2006av, Barnich:2010eb}.
As it turns out, the BMS$_3$ algebra can be obtained by taking a ultra-relativistic limit, known mathematically as a \.In\"on\"u-Wigner contraction of the infinite dimensional Virasoro algebra \cite{Barnich:2012aw}, a feature clearly special to the 3 dimensional case. Even more curiously the non-relativistic limit of the Virasoro algebra gives rise to the so-called Galilean Conformal Algebra \cite{Bagchi:2012cy}, which is isomorphic to the BMS$_3$ algebra. This feature fails to be true when one includes supersymmetry \cite{Banerjee:2015kcx,Banerjee:2016nio} and higher-spin fields \cite{Gonzalez:2013oaa,Afshar:2013vka} in the theory as problems of unitarity might arise \cite{Banerjee:2016nio,Lodato:2016alv}.\\
Supersymmetric extensions of these asymptotical symmetry algebras of supergravity solutions have been studied in detail \cite{Barnich:2014cwa, Banerjee:2016nio}. Specifically, the
$\cN=1$ \cite{Barnich:2014cwa}, $\cN=2$ \cite{Lodato:2016alv,Fuentealba:2017fck} and $\cN=4$ \cite{Banerjee:2017gzj}
BMS$_3$ were obtained by the conservative approach of direct asymptotic symmetry analysis of 3D flat solutions at null infinity. \\
In this paper we will use the conservative approach of \cite{Banerjee:2017gzj}, and derive the maximally $\cN$-extended super-BMS$_3$ algebra from a precise asymptotic symmetry analysis of the 3D solution at null infinity. The corresponding supergravity theory was first analysed in \cite{HOWE1996183}.  New feature of this asymptotic algebra are the presence of a non-abelian internal R-symmetry algebra and the fact that the spinors transform under a real and not necessarily irreducible representations thereof.
The results are, for the special case of $\cN=8$ super-BMS$_3$ (with SU$(2)$ R-symmetry), in agreement with the ultra-relativistic (unitary) democratic \.In\"on\"u-Wigner contraction of the corresponding super-Virasoro algebras presented in \cite{Banerjee:2016nio}. 
Similar constructions for asymptotic AdS supergravity solutions were presented in \cite{Henneaux:1999ib}, in which the supergenerators anticommutators were shown to close with quadratic non-linearities \cite{Fradkin:1992bz, Fradkin:1992km, Bowcock:1992bm}. As we shall see later on, this special feature persist in our analysis (see \eqref{BMS}).

Clearly, the non-linearities in the supersymmetry algebra have physical
implications such as the raising of the lower bound of
supersymmetric states energy. As usual, the bound is saturated by Minkowski space which is a ground state of the theory with trivial holonomy but other solutions whose global structure is non-trivial do exists. These are non-supersymmetric cosmological solutions, satisfying the energy bound. We will study their geodesics preoperties and give their classical entropy. Our analysis follows the one performed in \cite{Fuentealba:2017fck} for $\cN=2$ super BMS$_3$. \\
The paper is organized as follows: in section \ref{sec1}, we review the main features of $3$-dimensional (super)gravity theories. In section \ref{sec2}, we present the maximal $\cN$-extended non-linear super BMS$_3$ algebra. This is one of the main results of this paper. From the supercharges anticommutators, we derive the energy bounds focusing only on the the Neveu-Schwarz sector (anti-periodic boundary conditions for the spinors). We end this section by giving the asymptotic killing spinors of the solution at null infinity which parametrize the asymptotic supersymmetry invariance of the bosonic solution. In section \ref{sec3} we study the conditions imposed by holonomy and find the most generic constant bosonic solution satisfying the energy bound imposed the asymptotic symmetry. We find classes of so-called cosmological solutions and analyze their thermodynamic properties. We conclude the paper with some discussions and open issues in section \ref{sec4}. In first two appendices we will list our conventions (\ref{AppA}) and derive in detail aspects of the $\cN=8$ Super-BMS$_3$ algebra when the R-symmetry group coincides with SU$(2)$ (\ref{AppB}). The last two appendices (\ref{AppC},\ref{AppD}) are integral parts of this paper as they contain algebraic details that are suppressed for convenience in the main body of the paper. 

\section{ Supergravity in  $3$ dimensions  }\label{sec1}

Although 3D (super)gravity possesses the same conceptual hurdles of higher dimensional gravity theories and has no local bulk dynamical degree of
freedom (dof), it is still a perfect laboratory to approach quantum gravity, because it is renormalizable. There are three different
classical approaches to find these dof, namely : 1) Geometric
Structures,  2) The ADM formalism and 3) The Chern-Simons Formalism. We recommend interested readers to look at \cite{Carlip:1995zj,Carlip:2004ba} and references therein for further details on these approaches. The Chern-Simons Formalism of gravity was first introduced in \cite{Achucarro:1987vz}.
In this paper, we consider the Chern-Simons formulation of 3D gravity \cite{WITTEN198846}, that we discuss briefly in this section to make this work self-contained. The reader familiar with this formulation can skip to section \ref{sec2}. 

\subsection {Chern-Simons Formulation for $3$ dimensional
  gravity} \label{sec1.1}    
The Chern-Simons (CS) action on a three dimensional manifold $M$, invariant under the action of a compact Lie group G, is given by: 
\begin{equation}\label{csaction}
I [\cA] = \frac{k}{4 \pi}\int_M \langle \cA, d\cA+ \frac{2}{3}\cA^2 \rangle \;.
\end{equation} 
 Here the gauge field $\cA$ is regarded as a Lie-algebra-valued one form,
 and $\langle, \rangle $ represents a non-degenerate invariant
 bilinear form taking values on the Lie algebra space and acting as a metric and $k$ is level for the theory. Thus in a particular basis $\{T_a\}$ of the
 Lie-algebra, we can express $\cA= \cA^a_{\mu}\, T_a\, {\rm d}x^{\mu}$. The equation of motion is simply $$F \equiv d A + A \wedge A =0 .$$
The general solution of the equation of motion is topological, i.e. pure gauge. 
Consider for instance the Poincar\'e group $G={\rm ISO}(2,1)$ and a manifold $M$ with a {\it boundary}. The non-zero commutation relations of the Lie-algebra are:
\begin{equation}
[\cJ_a,\cJ_b]  = \epsilon_{abc}\cJ^c\,,\qquad [\cJ_a, P_b]  =
  \epsilon_{abc}P^c,
\end{equation}
where $a=1,2,3$ and $\epsilon_{abc}$ is the antisymmetric $3$-form. The explicit form of the gauge field is given in this basis by $\cA_{\mu}=
e^a_{\mu}P_a+ \omega^a_{\mu}J_a$, where $e^a_{\mu}$ acts as the vierbein and $\omega^a_{\mu}$ is the corresponding spin connection. The above action \eqref{csaction} then corresponds to the 3D Einstein-Hilbert action
 $$ S=\frac{1}{16\pi G}\int 2 e^a\,R_a \;, \quad R^a={\rm
  d}\o^a+\tfrac12\,\varepsilon^a{}_{bc}\,\o^b\,\o^c \;,$$
up to identifying the level $k=\frac{1}{4 G}$. Thus $3$-dimensional gravity invariant under the local ISO$(2,1)$ Poincar\'e group, with zero (or non-zero) cosmological constant can be cast as a 3-dimensional CS gauge theory with the
same gauge group. Indeed one can show that a generic ISO$(2,1)$ gauge
transformation parametrized by the element $U=E^aP_a+ \Omega^aJ_a$, act on the gauge field as
\begin{equation}
\delta \cA_{\mu}= - D_{\mu} U = -(\partial_{\mu} U + [\cA_{\mu}, U]).
\end{equation}
In terms of the gravity fields $(e^a_{\mu}, \omega^a_{\mu})$ the gauge transformation reads:
\begin{align}\label{GT}
\delta e_{\mu}^a&=-\partial_{\mu}E^a-\epsilon^{abc}e_{\mu b} \Omega_{c} - \epsilon^{abc} \omega_{\mu b} E_{c}\\
\delta \omega_{\mu}^a &= -\partial_{\mu} \Omega^a - \epsilon^{abc} \omega_{\mu b} \Omega_{c}
\end{align}
which are the expected local Lorentz transformations
generated by $\Omega^a$ and local diffeomorphism transformations generated by $E^a$. Recall that under a generic diffeomorphism transformation $x^{\mu} \rightarrow x^{\mu} +
V^{\mu},$ the fields $(e^a_{\mu}, \omega^a_{\mu})$ transforms as
\begin{align}\label{DT}
\tilde{\delta}e_{\mu}^a&= V^{\nu} (\partial_{\nu} e_{\mu}^a - \partial_{\mu} e_{\nu}^a)+ \partial_{\mu}(V^{\nu}e_{\nu}^a), \quad
\tilde{\delta}\omega_{\mu}^a= V^{\nu} (\partial_{\nu} \omega_{\mu}^a - \partial_{\mu} \omega_{\nu}^a)+ \partial_{\mu}(V^{\nu}\omega_{\nu}^a).
\end{align}
Thus for
$E^a= e^a_{\mu} V^{\mu}$ and turning off the local Lorentz transformation, we can show that the difference between
\eqref{GT} and \eqref{DT} is: 
\begin{align}
\tilde{\delta}e_{\mu}^a - \delta e_{\mu}^a = V^{\nu} (D_{\nu} e_{\mu}^a - D_{\mu}e_{\nu}^a) - \epsilon^{abc} V^{\nu} \omega_{\nu b} e_{\mu c}\;.
\end{align}
The 1st term of the RHS of the above equation, the torsion, vanishes
on-shell, while the 2nd term can be
identified with a local Lorentz transformation with parameter
$\Omega^a=\omega^a_{\mu}V^{\mu}$ \cite{WITTEN198846}. Thus we see that, on-shell, gauge transformation of Chern Simons theory is identical to local Lorentz and diffeomorphism transformation of $3$D Gravity.

We end this subsection by recalling how to find a nontrivial
classical solution in this theory.
Since \eqref{csaction} is a gauge theory, we first need to fix a
gauge. In
$(u,r,\phi)$ coordinates, for an arbitrary single valued group element
$U$, the general solution takes the form $A_{\mu}=
U^{-1}\partial_{\mu}U$. Imposing the gauge-fixing condition
$\partial_{\phi} A_{r}=0$, the connection will
have following form \cite{Banados:1994tn}:
\begin{equation}
A_r(r) = b(r)^{-1}\partial_r b(r), \quad A_{\phi}(r,\phi,u)= b(r)^{-1}A(\phi,u)b(r),
\end{equation}
where $b(r)$ and  $A(\phi,u)$ are arbitrary functions. To find $A_u$, we recall that the gauge fixing condition $\partial_{\phi} A_{r}=0$ must remain invariant under a new gauge transformation, for instance a time $(u)$ evolution, i.e. $\partial_u\partial_\phi \cA_u=0$. Using the equation of motion, this implies  $\partial_r\partial_{\phi}A_u=0$ which is solved generically by: 
\begin{equation}\label{Au0}
A_u (r,\phi,u) =  b(r)^{-1}B(\phi,u)b(r),
\end{equation}
where $B(\phi,u)$  is an arbitrary function of $\phi$ and time representing the residual gauge freedom of the system. Similarly $A(\phi,u)$ represents the residual part of the gauge field that can not be fixed. Instead, as we shall see in the
next subsection, they will give the global conserved charges and centrally
extended symmetry algebra at the boundary.  \\

Thus we see that, in a partial gauge fixed CS theory  the solution
will have the form $\cA=b(r)^{-1}(a+d)b(r),$ with $a= a_u du +a_{\phi} d
\phi$ is a function of $\phi$ and time. In the following, we choose 
$b(r)= e^{\alpha r}, \alpha$ a Lie-algebra valued constant, as a proper boundary condition on the field.

\subsection{Construction of Asymptotic symmetry algebra}
\label{sec1.2}

Once a solution of CS theory is obtained, one
can follow the canonical Hamiltonian approach of 
\cite{1974AnPhy..88..286R} to construct the conserved charges that correspond to the
residual global part of the gauge symmetry. Here, we shall only outline the procedure detailed in the original paper and \cite{Banados:1994tn}. \\

Consider a Chern-Simons theory on a manifold $\Sigma \times \mathbb{R}$, where $\Sigma$ is a compact two manifold and time is along $\mathbb{R}$. In this gauge theory, one defines global charges by demanding the differentials of the generators of gauge transformations to be regular for a certain choice of boundary conditions. Thus for some arbitrary
gauge transformation parameters $\lambda_a$ (matrix valued function) the charge needs to satisfy:
\begin{equation}\label{DQ}
\delta Q (\lambda)= -\frac{k}{2 \pi} \int_{\partial
  \Sigma}\lambda_a \delta A_{\mu}^a dx^{\mu} .
\end{equation}
Considering the example of ordinary $3$
dimensional gravity that we studied in the last section in
$(u,r,\phi)$ coordinate, the boundary consists of $\phi$
direction. Thus for this case, we get 
\begin{equation}\label{dQ}
\delta Q (\lambda)= -\frac{k}{2 \pi} \int_{\partial
  \Sigma}\lambda_a(\phi) \delta A^a_\phi d \phi\;.
\end{equation}
The above equation can be easily integrated to find the charges if 
 the parameter function  $\lambda$ is  independent
of the gauge field that are varied at the boundary. But in a generic cases, as the one studied in this paper, the gauge transformation parameter need not be completely independent of the gauge field. In such a scenario, one needs to explicitly find $\lambda$  using gauge variation equation in terms of independent parameters and gauge field components. The variations of the independent gauge parameter can be set to zero at the boundary and thus the above equation can be integrated to get the charge $Q(\lambda)$.  Our analysis (algebraic details in \ref{AppB}, \ref{AppC},  \ref{AppD}) follows the above procedure for computing $Q(\lambda)$
where the integration constant has been set to zero. It is clear from 
above that  expression for $Q(\lambda)$ depends the boundary value
of the gauge field $A^a(\phi)$.  \\
As we have seen in the last section, $A^a(\phi)$ is the residual part of
the gauge field that remains unfixed after gauge fixing. Similarly
$\lambda_a(\phi)$ corresponds to the residual part of the gauge
transformation parameters. Thus, we have constructed global 
charges that corresponds to the residual gauge symmetry . Expanding the
boundary fields and the parameters in modes, one can find find the
centrally extended algebra realized by this symmetry. We shall use
this technique in the next section to construct maximal $\cN-$extended
super BMS$_3$ algebra.

\section{Maximal $\cN-$Extended Super-BMS$_3$ algebra with nonlinear extension}\label{sec2}

In this section, we present the maximal $\cN-$ extended super-BMS$_3$ algebra. The maximally supersymmetric gravity theory under consideration contains one graviton $e_\mu{}^a$, eight (independent) gravitinos among $\psi_\alpha^{1,2}$ (see below for the range of $\alpha$), a set of R-symmetry gauge fields $\rho^a$ and a set of internal gauge field $\tilde{\phi}^a$. The theory is invariant under the super-Poincar\'e algebra:
\begin{align}
\label{SP} 
[J_n,J_m]=&(n-m)J_{n+m}\;, \qquad\quad\; [J_n,M_m]=(n-m)M_{n+m}\;, & 
\nonumber \\
 [\mathcal{R}^a,\mathcal{R}^b]=&{\rm i}\, f^{abc}\mathcal{R}^c\;, \qquad \qquad\qquad [\mathcal{R}^a,\mathcal{S}^b]={\rm i}\,f^{abc}\mathcal{S}^c & 
 \nonumber  \\
 [J_n,r^{(1,2),\alpha}_{p}]=&\left( \frac{n}{2}-p \right)r^{(1,2),\alpha}_{n+p},  \quad\qquad [\mathcal{S}^a,\mathcal{S}^b]=[\mathcal{S}^a,r^{(1,2),\alpha}_p]=0 &
 \nonumber \\
 \{r^{1,\alpha}_p,r^{1,\beta}_q \}=& M_{p+q}\eta^{\alpha\beta}-\frac{i}{6\hat\alpha}(p-q)(\lambda^a)^{\alpha\beta}S^a_{p+q},\quad [\mathcal{R}^a,r^{1,\alpha}_p]= i (\lambda^a)^{\alpha}_{\beta}r^{1,\beta}_p, & 
 \nonumber\\ 
 \{r^{2,\alpha}_p,r^{2,\beta}_p\}=& M_{p+q}\eta^{\alpha\beta}+\frac{i}{6\hat\alpha}(p-q)(\lambda^a)^{\alpha\beta}S^a_{p+q},\quad [\mathcal{R}^a,r^{2,\alpha}_p]=- i (\lambda^a)^{\alpha}_{\beta}r^{2,\beta}_p, & 
 \end{align}
%\end{subequations}
In the above, $J_n,M_n$
denote the Poincar\'e generators, $m,n$ run over
$(0,1,-1)$. The fermionic generators $r^{1,\alpha}_p,r^{2,\alpha}_p, \quad p,q= \pm \frac{1}{2}$ transform under a spinor representation $R$ of the internal algebra $G$, generated by $\cR^a$ (these are  R-symmetry generators as fermions transforms under them) and $\cS^a$. Generically, we can write the former generators in a representation $R$ as $(\lambda^a)^{\alpha\beta}$, satisfying the same commutation rules, i.e. $[\lambda^a,\lambda^b ] = f^{abc} \lambda^c $, where $(\lambda^a)^{\alpha\beta}=-(\lambda^a)^{\beta\alpha}$, $f^{abc}$ are the fully antisymmetric structure constants of the G and the indices $a,b,..=1,\dots, D$ while $\alpha, \beta,..=1,\dots, d$ with $D={\rm dim}(G)$ and $d={\rm dim }(R_G)$.\\
The metric $\eta^{\alpha\beta}$ of $R$ can be used to raise and lower spinor indices while the trace of the basis elements can be
expressed in terms of the eigenvalue of the second Casimir $C_{\rho}$  in the representation $R$. Here $\hat\alpha=\frac{C_{\rho}}{3(d-1)}$ is a constant. 
 This is the maximal $\cN-$extended super-Poincar\'e algebra in $3$ dimensions.\\
In the next section we start from a generic asymptotic gauge field and find the fall-off conditions which are consistent with the maximal $\cN-$extended asymptotic symmetry algebra. The required non-zero supertrace elements will have the following form ,
\begin{equation}\label{ST}
\langle J_m,M_n \rangle = \gamma_{mn}\;,\qquad \langle r^\alpha_-, r^\beta_+ \rangle= -\langle r^\alpha_+, r^\beta_- \rangle= 2 \eta^{\alpha\beta}\;,\qquad\langle\mathcal{R}^a,\mathcal{S}^b \rangle = \frac{4 C_\rho}{d-1}\delta^{ab}.
\end{equation}

\subsection{Super-BMS Algebra:}\label{sec2.2}

We work in the usual BMS gauge using Eddington-Finkelstein 
coordinates $(u,r,\phi)$. The Chern-Simons gauge field can then be written in the basis of the  global algebra generators as:  
\begin{align}\label{GF}
\mathcal{A} &=\left(M_1-\frac{1}{4}\mathcal{M}M_{-1} +\frac{1}{24\hat\alpha}\rho^a S^a  \right)du+\frac{dr}{2}M_{-1} 
\nonumber\\ 
+&\left(
   J_1+rM_0-\frac{1}{4}\mathcal{M}J_{-1}-\frac{1}{4}\mathcal{N}M_{-1}
   + \mathfrak{A} \psi_{\alpha}^1 r^{-,\alpha}_1    - 
\bar{\mathfrak{A}} \psi_{\alpha}^2 r^{2, -}_{\alpha}
   +\frac{1}{24\hat\alpha}\rho^a \mathcal{R}^a+
\frac{1}{24\hat\alpha}\tilde{\phi}^a \mathcal{S}^a \right)d\phi\;.
\end{align}
The various fields ${\mathcal{M},\mathcal{N},\rho^a,\psi_{\alpha}^1,\psi_{\alpha}^2,\tilde{\phi}^a}$ will only depend on $u$ and $\phi$ at null infinity and:
$$ \hat\alpha=\frac{C_{\rho}}{3(d-1)}\;,\qquad \bar{\mathfrak{A}}^2=
\mathfrak{A}^2=-1/4\;.$$
  It is easy to see that the above gauge field encodes the asymptotic flat metric :  
\begin{align}\label{PBM}
{\rm d}s^2=\gamma_{nm}e^ne^m=\mathcal{M}{\rm d}u^2-2{\rm d}u{\rm d}r+\mathcal{N}{\rm d}u{\rm d}\phi+r^2{\rm d}\phi^2 ,
\end{align}
where $\gamma_{nm}$ is the induced metric
\footnote{We can calculate the vierbeins as the coefficients of the translation generators :
\begin{align*}
e^{-1}=-\frac{1}{4}\mathcal{M}{\rm d}u-\frac{1}{4}\mathcal{N}{\rm d}\phi+\frac{1}{2}{\rm d}r
        \;, \quad 
e^0=r {\rm d}\phi \;, \quad
e^1={\rm d}u \;.
\end{align*}} on this space : $\gamma_{00}=1,~~\gamma_{1,-1}=-2.$ It is obvious that the above solution is globally different from 
Minkowski solution \footnote{The Minkowski metric in null coordinates is:  $ \quad ds^2=-{\rm d}u^2-2{\rm d}u{\rm d}r+r^2{\rm d}\phi^2 \;.$}.

Finally choosing the gauge: $\mathcal{A}=b^{-1}(a+{\rm d})b$ where $b=e^{\frac{r}{2}M_{-1}}$, the components of the gauge field $a$ read:
\begin{align}
\label{eq:auaphi}
a_u&=M_1-\frac{1}{4}\mathcal{M}M_{-1}+\frac{1}{24\hat\alpha}\rho^aS^a \;,
\nonumber\\
a_{\phi}&=J_1-\frac{1}{4}\mathcal{M}J_{-1}-\frac{1}{4}\mathcal{N}M_{-1}+\mathfrak{A}\psi^1_{\alpha}r^{-,\alpha}_1-\bar{\mathfrak{A}}\psi^2_{\alpha}r_2^{-,\alpha} +\frac{1}{24\hat\alpha}\rho^a \mathcal{R}^a+\frac{1}{24\hat\alpha} \tilde{\phi}^a_0 \mathcal{S}^a\;.
\end{align}
Various fields appearing above are constrained by gauge equation of motion $F=0$ \cite{Banerjee:2017gzj}. Next, we want to compute the gauge variation of this asymptotic field, generated by the most general gauge parameter:
\begin{align}\label{GGpara}
\Lambda=\xi^n_0 M_n+\Upsilon^nJ_n+\tilde{\lambda}^a_{S0}\mathcal{S}^a + \tilde{\lambda}^a_{R}\mathcal{R}^a + \zeta^{1,\alpha}_{\pm} r^{1,\alpha}_{\pm} + \zeta^{2,\alpha}_{\pm} r^{2,\alpha}_{\pm}\;,
\end{align}
where $\xi^n_0,\Upsilon^n,\tilde{\lambda}^a_{S0},\tilde{\lambda}^a_{R}$ are in general scalar functions of $(u,\phi)$ at null infinity. The gauge field and the parameter further need to satisfy the gauge variation equation,
\begin{equation}
\label{eq:deltaa}
\delta a = d\Lambda+[a,\Lambda].
\end{equation}
Now to compute the algebra, we first need to compute the conserved charges defined in \eqref{dQ}. To further integrate this relation to get the charges, one needs to express $\Lambda$ interms of independent parameters (whose variation can be set to zero at boundary) and gauge field components.  This can be obtained from the $\phi$ and $u$ components of the above gauge variation equation given as\footnote{Alternatively, one could derive the correct fall-off conditions for the gauge field and transformation parameter by combining the computation on the two chiral copies of AdS$_3$ , similarly to what was done in \cite{Banerjee:2017gzj}. See appendix \ref{AppC}.}:
\begin{equation}
\label{eq:deltaaphi}
\delta a_{\phi} = d_{\phi} \Lambda+[a_{\phi},\Lambda],\; \quad \delta a_{u} = d_{u} \Lambda+[a_{u},\Lambda]
\end{equation}
Next using the supertraces \eqref{ST} and the definition \eqref{dQ}, we can compute the variation of asymptotic charges $\mathcal{Q}(\lambda)$ of a $3$D maximally supersymmetric asymptotically flat solution as,
\begin{align*}
\delta\mathcal{Q}(\lambda) =& -\frac{k}{4 \pi} \int [\xi^1 \delta\mathcal{M} + \Upsilon^1 \delta\mathfrak{J} + 2 \mathfrak{A}\eta^{\alpha \beta} \zeta^1_{+,\alpha} \delta\psi^1_{\beta} + 2 \mathfrak{\bar{A}}\eta^{\alpha \beta} \zeta^2_{+,\alpha} \delta\psi^2_{\beta} + \tilde{\lambda}_R^a \delta\rho_a+\tilde{\lambda}_S^a \delta\tilde{\phi_a}]\;.
\end{align*}

The gauge field components $\mathcal{M},\mathfrak{J},\rho^a,\tilde{\phi}^a,\psi^1_{\beta} ,\psi^2_{\beta} $ are independent functions of $\phi$ only. Similarly $\xi^1,\Upsilon^n,\tilde{\lambda}^a_{S},\tilde{\lambda}^a_{R},\zeta^1_{+,\alpha} ,\zeta^2_{+,\alpha} $ are only $\phi$ dependent gauge transformation parameters. These are independent of the gauge field and their boundary variations are considered as zero. With this condition, the above expression can be integrated to give the conserved charge $\mathcal{Q}(\lambda)$ as,
\begin{align*}\label{CC}
\mathcal{Q}(\lambda) =& -\frac{k}{4 \pi} \int [\xi^1\mathcal{M} + \Upsilon^1 \mathfrak{J} + 2 \mathfrak{A}\eta^{\alpha \beta} \zeta^1_{+,\alpha} \psi^1_{\beta} + 2 \mathfrak{\bar{A}}\eta^{\alpha \beta} \zeta^2_{+,\alpha} \psi^2_{\beta} + \tilde{\lambda}_R^a \rho_a+\tilde{\lambda}_S^a \tilde{\phi_a}]\;.
\end{align*}
Finally we derive the asymptotic algebra by using the relation
\begin{equation*}
\{\mathcal{Q}[\lambda_1],\mathcal{Q}[\lambda_2]\}_{PB}=\delta_{\lambda_1}\mathcal{Q}[\lambda_2]\;,
\end{equation*} 

where the variation of the charge follows from the above expression.
The non-zero Poisson Brackets between the Fourier modes of the charges are:
\begin{align}\label{UBMS}
\{\mathfrak{J}_n,\mathfrak{J}_m\}&={\rm i}(n-m) \mathfrak{J}_{n+m}\;,  \qquad\qquad\;\;
\{\mathfrak{J}_n,\mathfrak{M}_m\}={\rm i}(n-m)\mathfrak{M}_{n+m}+{\rm i}\frac{c_M}{12}n^3\delta_{n+m,0}\;,
\nonumber\\
\{R^a_n,R^b_m\}&=-f^{abc}R^c_{n+m}\;,\qquad\qquad\qquad  \{R^a_n,S^b_m\}={\rm i}\,n\hat\alpha c_M\delta^{ab}\delta_{n+m,0}-f^{abc}S^c_{n+m} \;,
 \nonumber\\
\{\mathfrak{J}_n,\psi^{(1,2),\alpha}_m\}&={\rm i}\left(\frac{n}{2}-m \right)\psi^{(1,2),\alpha}_{n+m}+\frac{k_l}{2k_B} (\lambda^a)^{\beta\alpha}(\psi^{(1,2),\beta}S^a)_{n+m}\;,
\nonumber\\
\{R^a_n,\psi^{1,\alpha}_p\}&=-  (\lambda^a)^{\alpha}_{\beta}\psi^{1,\beta}_{n+p}\;, \qquad\qquad\qquad\;\; \{R^a_n,\psi^{2,\alpha}_p\}= (\lambda^a)^{\alpha}_{\beta}\psi^{2,\beta}_{n+p}\;,
\nonumber \\
\{\psi^{1,\alpha}_n,\psi^{1,\beta}_m\}&=\frac{c_M}{6}n^2\delta_{n+m,0}\eta^{\alpha\beta}+\mathfrak{M}_{n+m}\eta^{\alpha\beta}-\frac{{\rm i}}{6\hat\alpha}(n-m)(\lambda^a)^{\alpha\beta}S^a_{n+m}  
\nonumber \\
&\quad-\frac{1}{144\hat\alpha^2}\{\lambda^a,\lambda^b\}^{\alpha\beta}\frac{1}{4}(S^aS^b)_{n+m}\;,  
 \nonumber \\
\{\psi^{2,\alpha}_n,\psi^{2,\beta}_m\}&=\frac{c_M}{6}n^2\delta_{n+m,0}\eta^{\alpha\beta}+\mathfrak{M}_{n+m}\eta^{\alpha\beta}+\frac{{\rm i}}{6\hat\alpha}(n-m)(\lambda^a)^{\alpha\beta}S^a_{n+m}  
\nonumber \\
&\quad-\frac{1}{144\hat\alpha^2}\{\lambda^a,\lambda^b\}^{\alpha\beta}\frac{1}{4}(S^aS^b)_{n+m}  \;.
\end{align}
Here $c_M=12 k=4/G_{\rm N}\;, k_l = k \, l$ where $l$ is the AdS radius that needs to be sent to infinity $l \rightarrow \infty$. Finally $ k_B= \frac{2 k_l C_\rho}{d-1}$ and the modes are given by:
\begin{align*}
\mathfrak{J}_n &= \frac{k}{4\pi}\int {\rm d}\phi e^{in\phi}\mathcal{J}\;, \quad
\mathfrak{M}_n = \frac{k}{4\pi}\int {\rm d}\phi e^{in\phi}\mathcal{M}\;, \quad
\psi^{1,\alpha}_n = \frac{k}{4\pi}\int {\rm d}\phi e^{in\phi} \psi^{1,\alpha}\;, \nonumber \\
\psi^{2,\alpha}_n &= \frac{k}{4\pi}\int {\rm d}\phi e^{in\phi} \psi^{2,\alpha}\;,\;\;
S^a_n = \frac{k}{4\pi}\int {\rm d}\phi e^{in\phi} \rho^a\;,\qquad 
R^a_n = \frac{k}{4\pi}\int {\rm d}\phi e^{in\phi} \tilde\phi^a\;.
\end{align*}
 Here $\mathfrak{J}_n$ almost behaves as the mode of the boundary stress tensor and act as spin two generators. Thus, every other fields (and hence their modes) should transform as a primary with proper weight under $\mathfrak{J}_n$. For example $S^a_m,R^a_m$ should transform as a primary of weight 1,   $\psi^a_m$ should transform as a primary of weight $3/2$ and $\mathfrak{M}_n$ should transform as a primary of weight 2. But, as we see the $\{\mathfrak{J}_n,\psi^a_m\}$ Poisson bracket contains an extra non-linear term while the $\{\mathfrak{J}_n,S^a_m\}, \{\mathfrak{J}_n,R^a_m\}$ Poisson brackets are zero. Thus, we can not treat $\mathfrak{J}_n$ as the proper mode of the boundary stress tensor. The resolution to this issue is well known. The proper stress tensor modes are obtained by adding quadratic  Sugawara-like terms to the modes $\mathfrak{J}_n$. Accordingly, the modes $\mathfrak{M}_n$ also need to be shifted  (see \cite{DiFrancesco:639405}). The Sugawara-like shifts read:
\begin{equation}
\mathfrak{J}_n\rightarrow \hat{\mathfrak{J}}_n=\mathfrak{J}_n+\frac{1}{24\hat\alpha}(R^aS^a)_{n}\;, \qquad
\mathfrak{M}_n\rightarrow \hat{\mathfrak{M}}_n=\mathfrak{M}_n+\frac{1}{48\hat\alpha}(S^aS^a)_{n}\;.
\end{equation}
The new modes satisfy the following algebra
\footnote{ We obtain the quantum algebra  from the classic Poisson Brackets by using the standard conventions: 
\begin{align*}
\{A_n,B_m\}_{PB}&=i[A_n,B_m]\;, \quad
\{A_n,B_m\}_{PB}=\{A_n,B_m\}\;.
\end{align*}},
\begin{align}\label{BMS}
[\hat{\mathfrak{J}}_n,\hat{\mathfrak{J}}_m]&=(n-m)\hat{\mathfrak{J}}_{n+m}+\frac{c_J}{12}n^3\delta_{n+m,0}\;, \quad
[\hat{\mathfrak{J}}_n,\hat{\mathfrak{M}}_m]=(n-m)\hat{\mathfrak{M}}_{n+m}+\frac{c_M}{12}n^3\delta_{n+m,0} \;,
\nonumber \\
[\hat{\mathfrak{J}}_n,\psi^{(1,2),\alpha}_{m}]&=\left(\frac{n}{2}-m \right)\psi^{(1,2),\alpha}_{n+m}\;, \qquad 
[\hat{\mathfrak{J}}_n,R^a_m]=-mR^a_{n+m}\;, \qquad\quad
[\hat{\mathfrak{J}}_n,S^a_m]=-mS^a_{n+m}\;, 
\nonumber \\
[R^a_n,R^b_m]&=n\,\hat\alpha\, c_R\delta^{ab}\delta_{n+m,0}+if^{abc}R^c_{n+m}\;,\quad
[R^a_n,S^b_m]=n\,\hat\alpha\, c_M\delta^{ab}\delta_{n+m,0}+if^{abc}S^c_{n+m}\;,
\nonumber \\
[R^a_n,r^{1,\alpha}_p]&= i (\lambda^a)^{\alpha}_{\beta}r^{1,\beta}_{n+p}\;, \qquad\qquad\qquad\qquad\;\; [R^a_n,r^{2,\alpha}_p]=- i (\lambda^a)^{\alpha}_{\beta}r^{2,\beta}_{n+p} \nonumber \\
\{\psi^{1,\alpha}_n,\psi^{1,\beta}_m\}&=\frac{c_M}{6}n^2\delta_{n+m,0}\eta^{\alpha\beta}+\hat{\mathfrak{M}}_{n+m}\eta^{\alpha\beta}-\frac{i}{6\hat\alpha}(n-m)(\lambda^a)^{\alpha\beta}S^a_{n+m}\nonumber \\
&-\frac{1}{48\hat\alpha}(S^aS^a)_{n+m}\eta^{\alpha\beta}       -\frac{1}{144\hat\alpha^2}\{\lambda^a,\lambda^b\}^{\alpha\beta}\frac{1}{4}(S^aS^b)_{n+m}\;,
 \nonumber \\
\{\psi^{2,\alpha}_n,\psi^{2,\beta}_m\}&=\frac{c_M}{6}n^2\delta_{n+m,0}\eta^{\alpha\beta}+\hat{\mathfrak{M}}_{n+m}\eta^{\alpha\beta}+\frac{i}{6\hat\alpha}(n-m)(\lambda^a)^{\alpha\beta}S^a_{n+m}\;,
\nonumber \\
&-\frac{1}{48\hat\alpha}(S^aS^a)_{n+m}\eta^{\alpha\beta}   -\frac{1}{144\hat\alpha^2}\{\lambda^a,\lambda^b\}^{\alpha\beta}\frac{1}{4}(S^aS^b)_{n+m}\;,
\end{align}
with other commutators being zero.
This is the most generic quantum maximal $\cN$-extended BMS$_3$. Here we have introduced two new central terms $c_J, c_R$ in the algebra, that are allowed by Jacobi identity \cite{Banerjee:2016nio}.   
We also notice that with respect to the modified $\hat{\mathfrak{J}}_n$, all the generators transform transforms appropriately, and the spurious non-linear term in the  $[\mathfrak{J}_n,\psi^a_m]$ commutator is also eliminated. However extra non-linear terms  quadratic in the $S^a$ generators still remain in the anti-commutators (see \cite{Henneaux:1999ib} for the corresponding superconformal algebras). Note that the non-linear terms are a manifestation of the generic choice of representation for the internal symmetries.\\ 
Earlier non-linear extension of the BMS$_3$ algebra were observed in \cite{Poojary:2017xgn}, but in that case they originated by allowing fluctuation in the conformal factor of the boundary metric. In our construction, the boundary metric is always fixed to Minkowski.\\
We end this section with a comment on a special case of $\cN=8$ super-BMS$_3$ algebra that was studied in \cite{Banerjee:2016nio}. In this case the internal gauge algebra was considered as $G=SU(2)$ and we choose the fundamental representation $F_G$, then $(\lambda^a) \sim \sigma^a$ with $\sigma^a$ Pauli matrices satisfying $\{\sigma^a,\sigma^b\}=2i\delta^{ab} I$ \footnote{$\sigma$'s are different from $\lambda$'s, as they are not antisymmetric.}. It can be seen that for this case, the non-linear terms in the anticommutators cancel (see \ref{AppB}). This result is consistent with the corresponding superconformal algebra \cite{Ito:1998vd}, that closes with out any non-linear corrections.
 
\subsection{BMS Energy Bound}
\label{sec2.2}
As it is well-know, supersymmetry imposes constraints on the energy of supersymmetric states. The bounds are directly obtained from the super algebra. If we focus only on the NS sector of anti-periodic boundary conditions for the fermions, the global part of the algebra consists of the following generators :
\begin{equation}\label{GG} 
(\hat{\mathfrak{J}}_m , \hat{\mathfrak{M}}_m , \psi^{1,\alpha}_{\pm \frac{1}{2}} \psi^{ 2, \beta}_{\pm \frac{1}{2}}, R^a, S^a), \quad
m = −1,0,1 \;, \quad \alpha, \beta = 1,\dots d \;,\quad a= 1,\dots D\;. 
\end{equation} 

Following \cite{Donnay:2015abr}, we consider all possible positive-definite combinations of the supercharges
\begin{align*}
\{\psi^{1,\alpha}_{\frac{1}{2}},\psi^{1,\beta}_{-\frac{1}{2}} \}      +         \{\psi^{1,\alpha}_{-\frac{1}{2}},\psi^{1,\beta}_{\frac{1}{2}} \}+\{\psi^{2,\alpha}_{\frac{1}{2}},\psi^{2,\beta}_{-\frac{1}{2}} \}+\{\psi^{2,\alpha}_{-\frac{1}{2}},\psi^{2,\beta}_{\frac{1}{2}} \} \geq 0 \;,
\end{align*}
which explicitly gives:
\begin{align}\label{EB}
\hat{\mathfrak{M}}_0\geq -\frac{c_M}{6}+\frac{1}{48\hat\alpha}(S^aS^b)_0\delta_{ab}+\frac{1}{156\hat\alpha^2}\{ \lambda^a,\lambda^b \}^{\alpha\beta}\eta_{\alpha\beta}(S^aS^b)_0 \geq- \frac{1}{8G}\;.
\end{align}
 As explained in \cite{Banerjee:2017gzj}, the correct bound is obtained by considering the Sugawara-shifted generators. Note that, because of the non-linear quadratic corrections the energy bound is raised, hence supersymmetric ground states must have a higher energy. Also as pointed in \cite{Henneaux:2015ywa} and shown in \cite{Fuentealba:2017fck}, there are infinite number of bounds coming for all possible modes of the fermionic generators and all of them needs to be satisfied for an unitary theory. But the bound reported in \ref{EB} is the strongest one for the anti-periodic boundary conditions on fermions and hence once this is satisfied we get unitarity. It is easy to see that Minkowski vacuum $\hat{\mathfrak{M}}_0 = \mathfrak{M}_0 = -\frac{1}{8G}$, with all other fields vanishing, still saturates the bound. In this $\cal{N}-$ extended case, it is possible to saturate the bound for appropriate representation matrix $\lambda^a$.\\
In this paper, we will use the above bound \ref{EB} to constrain the general solutions of $3$D supergravity.

\subsection{Asymptotic Killing Spinors}\label{sec2.3}

In order to find fully supersymmetry backgrounds one imposes the vanishing of all the fermions and their supersymmetry variations. Among those, the first constraint simply imposes the variations of all bosonic fields to zero at null infinity whereas the vanishing of the gravitino variation constitutes the Killing spinor equations, the solutions of which parametrize the fermionic isometries of the background. Since we are interested in \eqref{BMS} symmetry at null infinity, only point to appreciate is that, as we have seen in the previous section, we need to perform Sugawara shifts to certain generators to get the correct algebra. With this in hindsight,
 we begin with a modified gauge field component $a_{\phi}$, incorporating the Sugawara shifts in the gauge field itself, such that it produce the correct $BMS_3$ algebra \eqref{BMS}. It takes the following form,
\begin{align}\label{cAphi}
a_{\phi}&=J_1-\frac{1}{4}\left( \mathcal{M}-\frac{1}{48\hat\alpha}\rho^a\rho^a       \right)J_{-1} - \frac{1}{4}\left(      \mathcal{N}-\frac{1}{24\hat\alpha}\tilde{\phi}^a \rho^a\right)M_{-1} \nonumber \\
&\quad+\mathfrak{A}\psi^1_{\alpha}r^{-,\alpha}_1-\bar{\mathfrak{A}}\psi^2_{\alpha}r_2^{-,\alpha} +\frac{1}{24\hat\alpha}\rho^a \mathcal{R}^a+\frac{1}{24\hat\alpha} \tilde{\phi}^a_{0} \mathcal{S}^a\;.
\end{align}
By abuse of notation we have used same $\mathcal{M}$ in the above expression but we keep in mind that the modes of this field appear in the algebra \eqref{BMS}.
Next we now analyze the fermion variations to calculate the asymptotic Killing spinors. For $\psi^1_{\alpha}$ the variation takes the form:
\begin{align*}
	\mathfrak{A}\delta \psi^1_{\alpha}=&-(\zeta^1_{+,\alpha} )''+\mathfrak{A}\Upsilon^+(\psi_{\alpha})'+\frac{3\,}{2}\mathfrak{A}(\Upsilon^+)'\psi_{\alpha}+\frac{1}{12\hat\alpha}(\lambda^a)^{\beta}_{\alpha}\rho^a(\zeta^{+,\beta}_1)'+\frac{1}{24\hat\alpha}(\lambda^a)^{\beta}_{\alpha}(\rho^a)'\zeta^{+,\beta}_1 \\
	&+\frac{1}{4}\left( \mathcal{M}-\frac{1}{48\hat\alpha}\rho^a\rho^a       \right)\zeta^{\alpha}_{1,+}-\frac{\mathfrak{A}}{24\alpha}(\lambda^a)^{\beta}_{\alpha}\rho^a\Upsilon^+\psi_{\beta}+\mathfrak{A}\psi_{\beta}(\lambda^a)^{\beta}_{\alpha}\lambda^a_R-\frac{1}{8}\frac{1}{144\hat\alpha^2}\rho^a\rho^b\{\lambda^a,\lambda^b\}^{\delta}_{\alpha}\zeta^1_{+,\delta}\;.
\end{align*}
Similar expression holds for $\delta\psi^2_{\alpha}$. 
Setting all fermions to zero, the final variation equations for both gravitinos read: 
\begin{align*}
\mathfrak{A}\delta \psi^1_{\alpha}=	(\zeta^1_{+,\alpha})''-\frac{1}{12\hat\alpha}(\lambda^a)^{\beta}_{\alpha}\rho^a(\zeta_{+,\beta}^1)'-\frac{1}{4}\left( \mathcal{M}-\frac{1}{48\hat\alpha}\rho^a\rho^a       \right)\zeta^1_{+,\alpha}+\frac{1}{8}\frac{1}{144\hat\alpha^2}\rho^a\rho^b\{\lambda^a,\lambda^b\}^{\delta}_{\alpha}\zeta^1_{+,\delta} &=0\;, \nonumber \\
\bar{\mathfrak{A}}\delta \psi^2_{\alpha}	=(\zeta^2_{+,\alpha} )''-\frac{1}{12\hat\alpha}(\lambda^a)^{\beta}_{\alpha}\rho^a(\zeta^2_{+,\beta})' 
	-\frac{1}{4}\left( \mathcal{M}-\frac{1}{48\hat\alpha}\rho^a\rho^a       \right)\zeta^2_{+,\alpha}+\frac{1}{8}\frac{1}{144\hat\alpha^2}\rho^a\rho^b\{\lambda^a,\lambda^b\}^{\delta}_{\alpha}\zeta^2_{+,\delta}&=0\;.
\end{align*}
The solutions of the above differential equations are :

\begin{align}\label{AKS}
\zeta^1_{+,\alpha} =& \big( e^{\frac{1}{24\hat\alpha} \lambda_a \rho_a \phi} \big)_\alpha ^\beta  ~\bigg[ {c}_{1\beta}~ e^{\frac{\sqrt{\left( \mathcal{M}-\frac{1}{48\hat\alpha}\rho^a\rho^a       \right)}}{2} \phi} + {c}_{2\beta}~ e^{-\frac{\sqrt{\left( \mathcal{M}-\frac{1}{48\hat\alpha}\rho^a\rho^a       \right)}}{2} \phi}\bigg]\;,
 \nonumber \\
\zeta^2_{+,\alpha} =& \big( e^{\frac{1}{24\hat\alpha} \lambda_a \rho_a \phi} \big)_\alpha ^\beta  ~\bigg[    \tilde{c}_{1\beta}~ e^{\frac{\sqrt{\left( \mathcal{M}-\frac{1}{48\hat\alpha}\rho^a\rho^a       \right)}}{2} \phi} + \tilde{c}_{2\beta}~ e^{-\frac{\sqrt{\left( \mathcal{M}-\frac{1}{48\hat\alpha}\rho^a\rho^a       \right)}}{2} \phi}\bigg].
\end{align}
Here $ c_{i\beta}, \tilde  c_{i\beta}, (i=1,2)$ are four spinors that can in general be functions of $u$. The $u-$ derivative equation of the spinors can be computed using the $u-$ component of gauge variation equation \eqref{eq:deltaaphi}. For our choice of (Sugawara shifted) gauge field  as in \eqref{eq:auaphi}, the $u$ dependence is trivial and these are indeed constant spinors\footnote{A generic case has been noted in section \ref{sec3} footnote.}.
The solutions are consistent with the periodicity of $\phi$ only when $\mathcal{M}-\frac{1}{48\hat\alpha}\rho^a\rho^a = - n^2$ and $n$ a strictly positive integer and $\lambda_a \rho_a$ is imaginary or zero. These conditions are satisfied for Minkowski vacuum ($\rho^a=0,\mathcal{M}=-1$) which is a fully supersymmetric solution. For $n=0$, the solutions become degenerate and only half the supersymmetries are allowed.
\section{Generic Bosonic Solutions}\label{sec3}

 In this section, we shall explore a class of purely bosonic topological $3$D gravity solutions, with non-trivial holonomy \cite{Bagchi:2012yk,Barnich:2012aw}.
 These solutions, as we shall see, will be cosmological in nature \cite{Cornalba:2002fi,Cornalba:2003kd}. 
We shall be looking for the corresponding bosonic 
solutions in this theory endowed with maximal $\cN$-extended supersymmetry at the null infinity .
Furthermore we shall henceforth restrict our analysis to zero mode solutions, for which all dynamical fields are constants.\\
Since the asymptotic symmetries are governed by $a_{\phi}$ \eqref{cAphi}, we do not modify this field. 
Also, as we are looking for a pure bosonic solution,
we set all fermionic components of the gauge field \eqref{cAphi} to zero.
Thus:
\begin{align}\label{Atheta}
a_{\phi}&=J_1-\frac{1}{4}\left( \mathcal{M}-\frac{1}{48\hat\alpha}\rho^a\rho^a       \right)J_{-1} - \frac{1}{4}\left(      \mathcal{N}-\frac{1}{24\hat\alpha} \tilde{\phi}^a \rho^a \right)M_{-1}  +\frac{1}{24\hat\alpha}\rho^a \mathcal{R}^a +\frac{1}{24\hat\alpha}\tilde{\phi}^a_0 \mathcal{S}^a\;.
\end{align}
Notice that the above $a_{\phi}$ is modified from \eqref{eq:auaphi} by incorporating the Sugawara shifts in various compunents as required to get \eqref{BMS}. Similarly, we also need to suitably modify the gauge transformation parameter $\Lambda$ from \eqref{GGpara} that reproduces the right conserved charge corresponding to \eqref{BMS} via the gauge variation equation \eqref{eq:deltaaphi}. Starting with the most generic gauge parameter and 
with a bit of algebra (see appendix \ref{AppD} for algebraic details), it can be shown that the required gauge parameter has the following form
\begin{align}\label{Gpara}
\Lambda=&\xi^1M_1+\Upsilon^1J_1+\left(    \lambda^a_R+\frac{1}{24\hat\alpha} \Upsilon^1 \tilde{\phi}^a                        \right)\mathcal{R}^a   
+ \left(  \lambda^a_S+\frac{1}{24\hat\alpha}\Upsilon^1 \tilde{\phi}^a       +\frac{1}{24\hat\alpha}\xi^1  \rho^a \right) \mathcal{S}^a 
\nonumber\\
&-\frac{1}{4}\Upsilon^1 \left(    \mathcal{M}-\frac{1}{48\hat\alpha} \rho^a\rho^a                                                        \right)J_{-1} 
-\frac{1}{4} \left[      \Upsilon^1\left( \mathcal{N} -\frac{1}{24\hat\alpha}   \rho^a\tilde\phi^a   \right)    +\xi^1    \left(   \mathcal{M}-\frac{1}{48\hat\alpha} \rho^a\rho^a                                   \right)                       \right]      M_{-1}.                 
\end{align}
This is a suitably truncated
version of \eqref{FP} where we have omitted the fermionic part for simplicity. Further the gauge field independent componenets of the parameter $\xi^1,\Upsilon^1,\lambda^a_R,\lambda^a_R$ has been taken to be constant and their  boundary variations has been set to zero. \\

Now to present a complete stationary circular symmetric bosonic solution of this system endowed with a maximal $\cN$-extended asymptotic supersymmetry, we look at the time component $a_u$ of the CS gauge field. Few points to recall: 

\begin{itemize}
	\item to obtain the generic solution compatible with the asymptotic symmetry, we need to incorporate the chemical potentials into the system \cite{Henneaux:2013dra,Gary:2014ppa,Matulich:2014hea}, which give vacuum expectation value to the time component of the gauge field $a_u$. These potentials can also be thought of as Lagrange multiplier associated to the dynamical fields of the system defined as the coefficients of the lowest weight components of the symmetry algebra.
	
	\item as we have shown in section \ref{sec1.1}, the diffeomorphism transformation of gravity is equivalent to the gauge transformation of the CS gauge theory. Thus, the time evolution of the various dynamical components of $a_\phi$ is generated by a gauge transformation whose components are now given by the chemical potentials (or Lagrange multipliers). This readily implies \footnote{The gauge transformation of $a_{\phi}$ by gauge parameter $\Lambda(\mu)$ is : $$\delta_\mu a_\phi=d_\phi \Lambda(\mu) + [a_\phi, \Lambda(\mu)],$$ whereas its time evolution from the equation of motion takes the form:  
		$$ {\rm d}_u a_\phi = {\rm d}_\phi a_u + [a_\phi, a_u].$$These two are identical if $a_u \sim \Lambda(\mu)$.} that the $a_u$ will have a similar form as \eqref{Gpara},
\end{itemize}
\begin{align}\label{Au}
a_u=&\mu_M M_1+\mu_J J_1+\left(    \mu^a_R+\frac{1}{24\hat\alpha} \mu_J \rho^a       \right)\mathcal{R}^a   
+ \left(  \mu^a_S+\frac{1}{24\hat\alpha}\mu_J \tilde{\phi}^a       +\frac{1}{24\hat\alpha}\mu_M \rho^a \right) \mathcal{S}^a \nonumber \\
-\frac{1}{4}   \mu_J & \left(    \mathcal{M}-\frac{1}{48\hat\alpha} \rho^a \rho^a            \right)J_{-1} 
-\frac{1}{4} \left[      \mu_J \left( \mathcal{N} -\frac{1}{24\hat\alpha}   \tilde{\phi}^a \rho^a   \right)    +\mu_M    \left(   \mathcal{M}-\frac{1}{48\hat\alpha} \rho^a \rho^a \right)                       \right]      M_{-1}\;,                  
\end{align}

where ${\mu_J, \mu_M,\mu^a_S,\mu^a_R }$	are the chemical potentials and their boundary variations are taken to zero. We have only turned on the chemical potentials corresponding to bosonic lowest weight generators as we are interested in pure bosonic solution. This can certainly be generalised to more generic scenario.
\begin{itemize}
	\item finally the above solutions have to satisfy appropriate regularity constraints related to the holonomy. In particular, the regularity of the solution requires trivial holomony in presence of a contractible cycles $\cal C,$ i.e.  
	\begin{equation}\label{Holo}
	H_{\cal C}= P e^{\int_{\cal C} a_{\mu} dx^{\mu}} = \pm I \;.
	\end{equation}
For the theory under consideration defined on a $3$D manifold $\Sigma \times \mathbb{R}$ we only require the holonomy along time direction to be trivial, i.e. the above condition \eqref{Holo} must be satisfied for the time component of the gauge field $a_u$. \\ 
\end{itemize}

Once the holonomy condition \eqref{Holo} and the energy bound as given in section \ref{sec2.3} is respected, we get a regular solution with required asymptotic falloff properties for our system.
One last important caveat to notice is that to solve the above holonomy constraint one needs an explicit matrix representation of the symmetry generators, which in general is not known. However, as pointed out in \cite{Matulich:2014hea,Bunster:2014mua}, one can exploit the pure-gauge (topological) nature of the solutions to gauge away the components proportional to the supertranslation generators $M$ and internal generators $\cR^a$ and $\cS^a$, which do not have an explicit matrix representation. The new component of the gauge field will now depend only on the superrotations generators $J$ (see appendix \ref{AppA} for their explicit matrix representation) and can be used to impose explicitly the above holonomy condition.\\
 To do so, we choose the general gauge group element $g=e^{\lambda_0M_0}$, which transforms the gauge field component as:   
\begin{align}\label{Aug}
a_u^g&=g^{-1}a_ug=e^{-\lambda_0M_0}a_ue^{\lambda_0M_0} \nonumber  \\
&=a_u+\lambda_0\mu_J M_1+\frac{1}{4}\lambda_0 \left[     \mu_J \left(   \mathcal{M} - \frac{1}{48\hat\alpha}\rho^a\rho^a \right)             \right] M_{-1}\;,
\end{align}	  
where $a_u$ is given as in \eqref{Au}. Fixing $\lambda_0$ and the chemical potential to the values: 
\begin{align}\label{CP1}
\lambda_0& = -\frac{\mu_M}{\mu_J}\;, \quad   
\mu_M=-\frac{\mu_J}{2}\frac{ \left(    \mathcal{N}-\frac{1}{24\hat\alpha}\tilde{\phi}^a \rho^a                    \right)}{                 \left(         \mathcal{M}  -\frac{1}{48\hat\alpha}  \rho^a\rho^a \right)}\;,   \\
\mu^a_R &=-\frac{1}{24\hat\alpha} \mu_J \tilde{\phi}^a  -  \frac{1}{24\hat\alpha} \mu_M \rho^a\;, \quad \mu^a_S= -\frac{1}{24\hat\alpha} \mu_J \rho^a \;, 
\end{align}
the time component of the gauge field, now depends only on superrotations generators and hence matrix representable \footnote{
	Since our initial BMS solution of \eqref{GF} does not contain $J$ generators, the holonomy condition is trivially satisfied after above gauge fixing.}
\begin{align}\label{Augf}
a^g_u=&\mu_J J_1   
-\frac{1}{4}   \mu_J  \left(    \mathcal{M}-\frac{1}{48\hat\alpha} \rho^a\rho^a  .                    \right)J_{-1} \;,
\end{align} 
Finally we can impose the regularity of the solution. Specifically, the gauge field $a_{\tau} = i a^g_u$ can be diagonalised with eigenvalues 
\begin{align}\label{EV}
\omega =& \pm i \mu_J\sqrt{\frac{1}{4}\left( \mathcal{M} -\frac{1}{48\hat\alpha} \rho^a \rho^a  \right)}\;.
\end{align}
Now, in order for this to have a trivial holonomy $\omega =\pm i \pi m$ where $m \in \mathbb{Z}$.
~~~\\
This condition  fixes the chemical potential $\mu_J$ in terms of the fields and an arbitrary integer $m$ to be:
\begin{align}\label{CP}
|\mu_J| = \frac{2\pi m}{(\mathcal{M} -\frac{1}{48\hat\alpha} \rho^a \rho^a)^\frac{1}{2}}\;,
\end{align}
and by the above set of relations \eqref{CP1} and \eqref{CP}, all chemical potentials are now fixed in terms of the zero modes of the fields. Thus we obtain the generic 3D bosonic zero mode solution given by \eqref{Atheta} and \eqref{Au} in a gravity theory with maximal bulk supersymmetry \eqref{SP}   and maximal $\cN$-extended non-linear asymptotic supersymmetry \eqref{BMS}.  Since in our construction we have implicitly assumed $(\mathcal{M} -\frac{1}{48\hat\alpha} \rho^a \rho^a)>0$, the solution satisfies the energy bound \eqref{EB} but there exist no well defined asymptotic killing spinors \eqref{AKS}\footnote{In this background with constant bosonic chemical potentials, the $u$ dependence of the killing spinor will be non trivial as,
\begin{align*}
\dot{\zeta}^{1,\alpha}_{+} &= \mu_J  (\zeta^{1,\alpha}_{+})' - i (\lambda^a)^{\beta}_{\alpha} \mu^a_R \zeta^{1,\beta}_{+} .
\end{align*} }. Hence this class of partially gauge fixed solutions are non-supersymmetric and nontrivial only at the boundary. The space time geometry in Bondi coordinates reads:
\begin{align}\label{FullM}
{\rm d}s^2
=& ( M+r^2 \mu_J^2)  {\rm d}u^2- 2 \mu_M {\rm d}u{\rm d}r + (J+ 2r^2 \mu_J) {\rm d}u {\rm d}\phi + r^2 {\rm d}\phi^2 \;,
\end{align}
where,
\begin{align}\label{MJ}
M &= {\mu_M}  \left[   \mu_J \left( \mathcal{N} -\frac{1}{24\hat\alpha}   \tilde{\phi}^a \rho^a   \right)    +\mu_M    \left(   \mathcal{M}-\frac{1}{48\hat\alpha} \rho^a \rho^a \right)\right ]\;,\quad
J =\mu_M\left( \mathcal{N} -\frac{1}{24\hat\alpha}   \tilde{\phi}^a \rho^a  \right)\;.
\end{align}
The chemical potentials appearing in \eqref{FullM} are fixed as in \eqref{CP1} and \eqref{CP} with $m=1$ to avoid singularities in space-time. In particular, for $m=1$, $-\mu_M$ is the inverse Hawking temperature of the space time and $\mu_J$ is related to the chemical potential of the angular momentum $J$ of the system. As it is clear from \eqref{MJ}, for static configurations with $\mathcal{N}=0$, the system can have non-zero angular moment due to the presence of the internal gauge fields, a feature that was also observed in \cite{Fuentealba:2017fck}.

\subsection{Thermodynamics of the Solution}

So far we have presented the space time metric \eqref{FullM} in the usual Bondi coordinate. In this coordinate, the space time does not have any singularity. To understand the geometry better, following \cite{Barnich:2012xq}, let us rewrite the metric in Schwarzchild-like (ADM) coordinates as,
\begin{align}\label{ADM}
{\rm d}s^2 = -N^2 {\rm d}t^2+ \mu_M^2 N^{-2} {\rm d}r^2 + r^2 ({\rm d}\vartheta + N^{\vartheta} dt
)^2
\end{align}
where we define new coordinates as $t = u - f(r)$ and $\vartheta=\phi-g(r)$ and
$$
N^2 =  \frac{\tilde{A}^2}{4r^2} - \tilde{B}\;, \qquad
N^{\vartheta} = \frac{\tilde{A}}{2 r^2}\;.
$$
Here, with \eqref{MJ} we use compact notations $\tilde{A}$ as the coefficient of d$u$d$\phi$ and $\tilde{B}$ as the coefficient of d$u^2$ in the above metric \eqref{FullM}:
\begin{align}
\tilde{A} = J+2r^2 \mu_J  \;, \qquad
\tilde{B} = M+r^2 \mu_J^2\;.
\end{align}

Let us consider $\left(\mathcal{M} -\frac{1}{48\hat\alpha} \rho^a \rho^a\right) \geq 0$, hence a solution satisfying the energy bound \eqref{EB}. Under this condition \eqref{ADM} represents a cosmological spacetime .
In $(t,r,\vartheta)$ coordinate, the function $N^2$ vanishes at the hypersurface $r=r_c$, $(N^2)_{r=r_c} = 0$. This hypersurface is in fact a cosmological horizon and requiring $r_c> 0$ gives:
\begin{align}\label{CosH}
r_c = \frac{1}{2} \frac{| \mathcal{N} -\frac{1}{24\hat\alpha}   \tilde{\phi}^a \rho^a  |}{\left( \mathcal{M} -\frac{1}{48\hat\alpha} \rho^a \rho^a \right)^\frac{1}{2}}\;.
\end{align}
 To understand the nature of the horizon $r_c$, we write the above metric in a different coordinate system. For the region of the space time where $r>r_c$, let us define new coordinates $(T,X,\vartheta)$ as,  
 \begin{align}
 T^2= \frac{r^2-r_c^2}{\mathcal{M} -\frac{1}{48\hat\alpha} \rho^a \rho^a}\;, \qquad 
 X=\vartheta + \mu_J t\;.
 \end{align}
Similarly for the other region $r<r_c$, we define $(\hat T,X,\vartheta)$: 
  \begin{align}
 \hat T^2= \frac{r_c^2-r^2}{\mathcal{M} -\frac{1}{48\hat\alpha} \rho^a \rho^a}\;, \qquad 
 X=\vartheta + \mu_J t\;.
 \end{align}
In these coordinates, the space time metric is given by: 
\begin{align}\label{Torus}
{\rm d}s^2 &= -{\rm d}T^2 + \big(\mathcal{M} -\frac{1}{48\hat\alpha} \rho^a \rho^a\big)T^2 {\rm d}X^2 + r_c^2 {\rm d}\vartheta^2\;, \quad r>r_c \nonumber \\
&= {\rm d}\hat T^2 - \big(\mathcal{M} -\frac{1}{48\hat\alpha} \rho^a \rho^a\big)\hat T^2 {\rm d}X^2 + r_c^2 {\rm d}\vartheta^2\;, \quad r<r_c.
\end{align}
Thus in the outer region $r> r_c$, we have a cosmological space time of topology $R\times S^1 \times S^1$, a solid torus. Both $S^1$ factors have periodicity $2\pi$, the radius of the $\vartheta$ circle is fixed to $ r_c$, while the radius of the $X$ circle is $T$ dependent. It is also clear that, in the outer region we have closed space-like geodesics whereas in the inner region we can have closed time-like geodesics, as $X$ is a time-like coordinate in the interior. Thus, we readily conclude that $r=r_c$ is a Cauchy horizon \cite{Barnich:2012aw}. To avoid closed time-like curves, we cut the space-time at $r=r_c$. It can also be checked that $r=r_c$ is also a killing horizon. Finally, we can compute the Bekenstein-Hawking entropy associated with this class of Cauchy horizons:
\begin{align}\label{entropy}
S = \frac{2\pi r_c}{4G} = \frac{2\pi}{4G}\frac{1}{2} \frac{| \mathcal{N} -\frac{1}{24\hat\alpha}   \tilde{\phi}^a \rho^a  |}{\left( \mathcal{M} -\frac{1}{48\hat\alpha} \rho^a \rho^a \right)^\frac{1}{2}}
= \frac{\pi}{4G} \frac{| \mathcal{N} -\frac{1}{24\hat\alpha}   \tilde{\phi}^a \rho^a  |}{\left( \mathcal{M} -\frac{1}{48\hat\alpha} \rho^a \rho^a \right)^\frac{1}{2}}\;.
\end{align}
As expected, the entropy of the system is completely determined by the zero mode solution.
Alternatively, the entropy can be found using the Chern-Simons gauge field:
\begin{align}
S&=\frac{k}{2\pi} \int d\phi \langle a_u,a_{\phi}\rangle \nonumber \\
&=k \left[     \mu_J \mathcal{N} + \mu_M \mathcal{M} +\frac{1}{2}\tilde{\phi}^a \mu^a_S  + \frac{1}{2} \rho^a \mu^a_R             \right]
\end{align}
and plugging in the expressions \eqref{CP1}, \eqref{CP} for the chemical potentials, the entropy reduces to:
\begin{align}\label{ECS}
S = k \pi m \frac{| \mathcal{N} -\frac{1}{24 \hat\alpha} \tilde{\phi}^a \rho_a |}{\left( \mathcal{M} -\frac{1}{48\hat\alpha} \rho^a \rho^a  \right)^{\frac{1}{2}}}\;,
\end{align}
which matches with \eqref{entropy} for $m=1$. 
 The choice of $m=1$ sector is obvious, as only this sector is connected to the standard cosmological space time \eqref{FullM}.   

\section{Discussion and Outlook}\label{sec4}
With this paper we completed the detailed analysis of fall-off conditions necessary to obtain \emph{all} the supersymmetric extensions of the BMS algebras, presented in \cite{Banerjee:2016nio}. In the maximal $\cN$-extended super-BMS$_3$ case analyzed here we find non-linearity in the asymptotic algebra and modifications to the energy bounds for asymptotic states. Unlike $\cN=4,8$ super-BMS$_3$  studied respectively in \cite{Banerjee:2017gzj} and appendix \ref{AppB} of this paper, the non-linearity does not disappear after Sugawara-shifting the energy-momentum generators
%\footnote{For anti-periodic boundary conditions of fermionic generators, the non-linearity in energy bound as reported in %\cite{Fuentealba:2017fck} also disappears after proper modification of generators, as shown in \cite{Banerjee:2017gzj}.} . 
 Furthermore, we have shown that circular symmetric solutions that are flat cosmologies, satisfying $\left(\mathcal{M} -\frac{1}{48\hat\alpha} \rho^a \rho^a\right) > 0$,  are not supersymmetric. Similar results hold for abelian R-symmetry algebra as discussed in \cite{Fuentealba:2017fck}. There are three other distinct kinds of solutions \cite{Barnich:2012aw} that would appear for different conditions on the fields as presented below : \\
$a) \left(\mathcal{M} -\frac{1}{48\hat\alpha} \rho^a \rho^a\right) = 0\;$:  this class corresponds to \emph{null orbifold} solutions \cite{Horowitz:1990ap}. Here the asymptotic killing spinors \eqref{AKS} are degenerate and only half of them are independent. Hence this class of solution is only asymptotically half supersymmetric.\\
 $b)-\frac{1}{8G} <\left(\mathcal{M} -\frac{1}{48\hat\alpha} \rho^a \rho^a\right) < 0 \; $:  \emph{conical defect} solutions \cite{Deser:1983tn,Deser:1983nh}, satisfying the energy bound and asymptotically full supersymmetric.\\
 $c) \left(\mathcal{M} -\frac{1}{48\hat\alpha} \rho^a \rho^a\right) < -\frac{1}{8G} \;$: \emph{conical surplus} solutions that do not satisfy the energy bound.\\
 These solutions are not interesting from a cosmology perspective but are nevertheless non-trivial configurations of 3D gravity. Detailed discussions on the thermodynamics of their R-symmetry-abelian counterparts can be found in \cite{Fuentealba:2017fck} and references therein. For the non-abelian R-symmetry cases studied in this paper, most of the physics will be similar and hence we do not present the details here.\\
Let us end the paper with some interesting outlooks.
It is known that 3D gravity solutions with non-trivial topology correspond to stress-energy tensors  a two dimensional theory. It comes from the relation between a Chern Simons theory with a boundary and an associated chiral Wess-Zumino-Witten model \cite{Witten:1988hf,Moore:1989yh,Elitzur:1989nr}. As we have already seen, the non-trivial boundary for the Chern Simon theory (in our case the torus) comes from generic asymptotic fall off conditions on the gauge fields. It has been shown in \cite{Barnich:2013jla} for ordinary BMS$_3$ and in \cite{Barnich:2015sca} for ${\cal{N}}= 1 $ super-BMS$_3$ that one needs to add a suitable boundary term to the action for variation principle to go through. The fall off conditions also provide extra constraints to the Wess-Zumino-Witten model. Finding a similar two dimensional description for $\cal{N}$-extended super-BMS$_3$ obtained in this paper would provide a complete set of such $2-$dimensional theories that will act as dual to 3D asymptotically flat supergravity theories.

The second point is more generic and is related to the issue of understanding the implications of these infinite dimensional $3$-dimensional asymptotic symmetries on the dynamics of the corresponding two dimensional theory. As in $4$-dimensional gravity, we know
\cite{He:2014laa,Cachazo:2014fwa,Strominger:2014pwa} that the Ward identities of BMS$_4$ symmetries are related to bulk gravitational soft theorems. Interestingly, it has been very recently noticed by Barnich \cite{Barnich:2018zdg} that in $4$-dimensions there are also boundary degrees of freedom and they are highly constrained by BMS$_4$. In fact it has been proposed that the classical contribution to Bekenstein-Hawking entropy comes from these degrees of freedom. In the $3$-dimensional case,  there is no bulk graviton and hence we do not have a notion of soft theorem but the boundary theory and boundary degrees of freedom do exist. It would be interesting to study the of BMS$_3$ symmetry on their counting. Although the above issue is not directly related to study of this paper, but having (maximal)supersymmetry in the theory is technically helpful in counting the corresponding degrees of freedoms. We plan to report on this in future.

\vspace{1cm}
{\bf Acknowledgements}\\

We would like to thank Glenn Barnich, Suvankar Dutta, Diptimoy Ghosh, Dileep Jatkar, Wout Merbis, Sunil Mukhi and Arnab Rudra for useful discussions. N.B is thankful to ICTP for hospitality where part of the work was completed. We are also thankful to Keshav Dasgupta for pointing out a major typo is the first version of the paper. Our work is partially supported by the following Government of India Fellowships/Grants:
N.B. by a Ramanujan Fellowship, D.S.T. and T.N. by a UGC Fellowship. We thank the people of India for their generous support for the
basic sciences.

\appendix

\section{Conventions}\label{AppA}
In this paper we follow conventions similar to \cite{Lodato:2016alv}. \\
The antisymmetric Levi-Civita symbol has component $\epsilon_{012} =-\epsilon^{012}  +1$ and the tangent space metric is the 3D Minkowski metric
\begin{equation}
\eta_{ab} = \left( \begin{array}{ccc} -1 & 0 & 0 \\ 0 & 1 & 0 \\ 0 & 0 & 1\end{array}\;. \right)
\end{equation}
The $\Gamma$-matrices satisfying the three dimensional Clifford algebra $\{\Gamma_a, \Gamma_b\} =2 \eta_{ab}$ are:
\begin{equation}
\Gamma_0 =  {\rm i} \sigma_2 \,, \qquad \Gamma_1 = \sigma_1 \,, \qquad \Gamma_2 = \sigma_3\,,
\end{equation}
with $\sigma_i$ the Pauli matrices:
\begin{equation}
\sigma_1 = \left(\begin{array}{cc} 0 & 1 \\ 1 & 0\end{array} \right)\,, \qquad 
\sigma_2 = \left(\begin{array}{cc} 0 & -{\rm i} \\ {\rm i} & 0\end{array} \right)\,, \qquad
\sigma_1 = \left(\begin{array}{cc} 1 & 0 \\ 0 & -1\end{array} \right)\,.
\end{equation}
Finally, the charge conjugation matrix $C = {\rm i}\sigma_2$, or explicitly
\begin{equation}
C_{pm} = \varepsilon_{pm} = C^{pm}= \left(\begin{array}{cc} 0 & 1 \\ -1 & 0\end{array} \right)\,.
\end{equation}
Throughout this paper the fermionic indices $p,m$ run over $-,+$  (contrarily to \cite{Lodato:2016alv} where they run over $+,-$). The supercharges are also taken to be Grassmann quantities, as the fermion parameters and the gravitinos.\\ All spinors in this work are Majorana and the Majorana conjugate of a spinor $\psi^{\alpha}_p$ is $\bar{\psi}^{\alpha}_p = C_{pm}\psi^{\alpha m}$. Here $\alpha,\beta$ are internal indices. Our conventions imply that we can use the identities
\begin{align}
\Gamma_a\Gamma_b & = \epsilon_{abc}\Gamma^c + \eta_{ab} \mathbbm{1}\;, &&& \Gamma^a{}^{p}{}_{q} \Gamma_a{}^{r}{}_{s} & = 2 \delta^{p}_{s} \delta^{r}_{q} - \delta^{p}_{q}\delta^{r}_{s}\;, \\
C^T & = - C\,, &&& C \Gamma_a & = - (\Gamma_a)^T C\;.
\end{align}
In verifying the closure of the supersymmetry algebra on the fields and the off-shell invariance of the action, the three dimensional Fierz relation is useful.
\begin{equation}\label{Fierz}
\zeta\bar{\eta} = - \frac12 \bar{\eta}\, \zeta \, \mathbbm{1} - \frac12 (\bar{\eta}\Gamma^a \zeta)\Gamma_a\;,
\end{equation}
Other useful identities are:
\begin{equation*}
\bar\psi \G_a\,\eta=\bar\eta\,\G_a\,\psi\;,\qquad \bar\psi \G_a\,\epsilon= -\bar\epsilon\,\G_a\,\psi\;,
\end{equation*}
where $\psi,\eta$ are Grassmannian one-forms, while 
$\epsilon$ is a Grassmann parameter.
\noindent It is sometimes convenient to change basis of the tangent space to one more suited for the $sl(2,R)$   algebra in the bosonic sector of flat space supergravity. We do this by choosing a map to bring the generators of $SO(2,1)$ satisfying the commutator relations $[J_a,J_b] = \epsilon_{abc}J^c$) to those of $SL(2,\bR)$ satisfying $[L_n,L_m] = (n-m)L_{n+m}$. This defines a matrix $U^a{}_n$  such that: 
\begin{equation}
L_n = J_a\, U^a{}_n \,.
\end{equation}
An explicit representation of $U^a{}_n$ is for instance
\begin{equation}\label{Umat}
U^a{}_n = \left( \begin{array}{ccc} -1 & 0 & -1 \\ -1 & 0 & 1 \\ 0 & 1 & 0 \end{array} \right)\,.
\end{equation}
In this basis the tangent space metric $\eta_{ab}$ with $a,b=\{0,1,2\}$ is mapped to the metric $\gamma_{nm}$ defined below with $n,m= \{-1,0,+1\}$. The new gamma-matrices now satisfy a Clifford algebra with
\begin{equation}\label{gammadef}
\{\tilde{\Gamma}_m, \tilde{\Gamma}_n\} = 2 \gamma_{nm} \equiv 2 \left( \begin{array}{ccc} 0 & 0 & -2 \\ 0 & 1 & 0 \\ -2 & 0 & 0 \end{array}\right) \qquad \text{with } n,m= -1 , 0, +1\,.
\end{equation}
A real representation for the gamma matrices with $n,m$ indices can be obtained by taking $\tilde{\Gamma}_n= U^a{}_n \Gamma_a$, or explicitly:
\begin{align}
\tilde{\Gamma}_{-1} & =  - (\sigma_1 + {\rm i} \sigma_2) = \left(\begin{array}{cc} 0 & -2 \\ 0 & 0\end{array} \right)\,, \\ 
\tilde{\Gamma}_0 & = \sigma_3 = \left(\begin{array}{cc} 1 & 0 \\ 0 & -1\end{array} \right)\,, \\
\tilde{\Gamma}_{+1} & = \sigma_1 - i{\rm i} \sigma_2 = \left(\begin{array}{cc} 0 & 0 \\ 2 & 0\end{array} \right)\,.
\end{align}
In addition to the usual Clifford algebra, the gamma matrices now satisfy the commutation relations
\begin{equation}\label{Gammacom}
[\tilde{\Gamma}_{n} , \tilde{\Gamma}_{m}] = 2(n-m)\tilde{\Gamma}_{n+m}\,,
\end{equation}
which is exactly the $sl(2,\bR)$ algebra.

\section{$\cN=8$ Super-BMS$_3$}\label{AppB}

In this appendix, we demonstrate how the $N=8$ super-BMS$_3$ algebra does not get non-linear extension in the supercharges anticommutators. To do so, we prove this is the case for the asymptotic symmetry algebra for $3$D AdS gravity with $N=(4,4)$ supersymmetry. The gravitinos transform under the defining representation of the SU$(2)$ R-symmetry. The global super conformal algebra reads:

\begin{align*}
[L_n,L_m]&=(n-m)L_{n+m}\;, &
[R^i,R^j]&={\rm i} \epsilon^{ijk}\,R^k\;, \nonumber\\
[L_n,Q^{a\pm}_\a]&=\left(\frac{n}{2}-\alpha\right)Q^{a,\pm}_{n+\alpha}\;, & [L_n,R^i]&=0\;,\nonumber\\
[R^i,Q^{a\,+}_\a]&=-\tfrac12\,(\sigma^i)^a{}_b Q^{b+}_\a\;, &
[R^i,Q^{a\,-}_\a]&=+\tfrac12\,(\bar\sigma^i)^a{}_b Q^{b-}_\a\;,
\nonumber\\
\{Q^{a,\,+}_\a,Q^{b,\,-}_\b\}&=\delta^{ab}L_{\alpha+\beta}-(\alpha-\beta)\left(\sigma^i\right)^{ab}R^i\;, &
\{Q^{a,\pm}_\a,Q^{b,\pm}_\b\}&=0\;. & 
\end{align*}
The asymptotic gauge field we start from has the form:
\begin{equation}
A=\left(L_1+\frac{r}{l}L_0+\frac{r^2}{4l^2}L_{-1}-\frac{1}{2}\mathfrak{L}_+L_{-1}-\frac{1}{2}\psi_{a,+}Q^{a,+}_- +\frac{1}{2}\psi_{a,-}Q^{a,-}_- + i\phi^iR^i\right){\rm d}x^+  \nonumber
\end{equation}
Let us take the supertrace elements as
\begin{align*}
\langle L_n,L_m \rangle=\gamma_{nm}\;,~~~~~\langle Q^{a,+}_{\alpha},Q^{a,-}_{\beta}\rangle=\langle Q^{a,-}_{\alpha},Q^{a,+}_{\beta}\rangle =C_{\alpha\beta}\;,~~~~~\langle R_i,R_j\rangle= - \delta_{ij}\;.
\end{align*}
and the generic gauge parameter 
\begin{equation*}
\lambda=\chi^nL_n+\epsilon^{\alpha}_{a,+}Q^{a,+}_{\alpha}+\epsilon^{\alpha}_{a,-}Q^{a,-}_{\alpha}+\lambda^iR^i\;.
\end{equation*}
From the gauge variations, we first compute the constraint equations:
\begin{align*}
\chi^0&=-Y'+\frac{r}{l}Y\;, \\
\chi^-&=\frac{1}{2}Y''-\frac{r}{2l}Y'+\left(\frac{r^2}{4l^2}-\frac{1}{2}\mathfrak{L}_+\right)Y-\frac{1}{4}\sum_{a=1,2}\left(\psi_{a,+}\epsilon_{a,-}- \psi_{a,-}\epsilon_{a,+}\right)\;, \\
\epsilon^-_{a,+}&=-\epsilon'_{a,+}+\frac{r}{2l}\epsilon_{a,+}-\frac{1}{2} \psi_{a,+}Y +\frac{i}{2}\phi^i_R
\epsilon_{b,+}\left(\sigma^i\right)^b_a \;,\\
\epsilon^-_{a,-}&=-\epsilon'_{a,-}+\frac{r}{2l}\epsilon_{a,-}+\frac{1}{2} \psi_{a,-}Y -\frac{i}{2}\phi^i_R
\epsilon_{b,-}\left(\bar{\sigma}^i\right)^b_a\;.
\end{align*}
where $\epsilon^+_{a,\pm}=\epsilon_{a,\pm}$ and $\chi^+=Y$. \\

The other variation equations read: 
\begin{align*}
\delta\mathcal{L}_+=&-Y'''+2\mathcal{L}_+Y'+\mathcal{L}_+'Y+\frac{1}{2}\left(\psi_{a,+}'\epsilon_{a,-}+3\psi_{a,+}\epsilon_{a,-}'\right)-\frac{1}{2}\left(\psi_{a,-}'\epsilon_{a,+}+3\psi_{a,-}\epsilon_{a,+}'\right) \\
&+\frac{i}{2}\left[\psi_{a,+}\epsilon_{b,-}\phi^i\left(\bar{\sigma}^i\right)^b_a + \psi_{a,-}\epsilon_{b,+}\phi^i\left(\sigma^i\right)^b_a\right]\;, \\
\delta\psi_{a,+}=&2\epsilon_a''+ \left(\psi'_{a,+}Y+\frac{3}{2}\psi_{a,+}Y'\right) - i\left[{\phi^i}'\epsilon_{b,+}\left(\sigma^i\right)^b_a+2\phi^i\epsilon'_{b,+}\left(\sigma^i\right)^b_a\right]-\mathfrak{L}_+\epsilon_{a,+} \\
&-\frac{i}{2}\psi_{b,+}\phi^iY\left(\sigma^i\right)^b_a+\frac{1}{2} \lambda^i\psi_{b,+}\left(\sigma^i\right)^b_a-\frac{1}{2} \phi^i\phi^j\epsilon_{c,+}\left(\sigma^j\right)^c_b\left(\sigma^i\right)^b_a \;,\\
\delta\psi_{a,-}=&-2\epsilon_{a,-}''+ \left(\psi'_{a,-}Y+\frac{3}{2}\psi_{a,-}Y'\right)-i \left[{\phi^i}'\epsilon_{b,-}\left(\bar{\sigma}^i\right)^b_a+2\phi^i\epsilon'_{b,-}\left(\bar{\sigma}^i\right)^b_a\right]+\mathfrak{L}_+\epsilon_{a,-} \\
&+\frac{i}{2} \phi^i\left(\bar{\sigma}^i\right)^b_a\psi_{b,-}Y+\frac{1}{2} \phi^i\phi^j\epsilon_{c,-}\left(\bar{\sigma}^i\right)^b_a\left(\bar{\sigma}^j\right)^c_b-\frac{1}{2} \lambda^i\psi_{b,-}\left(\bar{\sigma}^i\right)^b_a\;, \\
i\delta\phi^i&={\lambda^i}'-\epsilon_{ijk}\phi^j\lambda^k+\frac{1}{2}\psi_{a,+}\epsilon_{b,-}\left(\sigma^i\right)^{ab}R_i+\frac{1}{2} \psi_{a,-}\epsilon_{b,+}\left(\sigma^i\right)^{ba}R_i\;.
\end{align*}
The charges are obtained from :
\begin{equation*}
\delta\mathcal{C}=-\frac{k}{4\pi}\int  {\rm d}\phi \langle\lambda ,\delta A_{\phi}\rangle\;.
\end{equation*}
Hence we get 
\begin{align*}
\mathcal{C}&=-\frac{k}{4\pi}\int {\rm d}\phi\left[\mathcal{L}_+Y+\frac{1}{2}\epsilon_{a,+}\psi_{a,-}-\frac{1}{2}\epsilon_{a,-}\psi_{a,+}-i\lambda_i\phi_i\right] \\
&=-\frac{2}{k}\left[\sum_{n} L_nY_{-n}+\sum_{\alpha}\frac{1}{2}\epsilon^{-\alpha}_{a,+}\hat{\psi}^{\alpha}_{a,+}-\sum_{\alpha}\frac{1}{2}\epsilon^{-\alpha}_{a,-}\hat{\psi}^{\alpha}_{a,-} -i \sum_{n} \lambda^{-n}_iR^n_i\right]
\end{align*}
We then derive the asymptotic algebra by using the relation
\begin{equation*}
\{\mathcal{C}[\lambda_1],\mathcal{C}[\lambda_2]\}_{PB}=\delta_{\lambda_1}\mathcal{C}[\lambda_2]
\end{equation*}
The Poisson brackets are
\begin{align*}
i\{L_n,L_m\}&= \frac{n^3k}{2}\delta_{n+m,0}+(n-m)L_{n+m,0}
\\
i\{L_n,\hat{\psi}^{a,+}_{\alpha}\}&= \left(\frac{n}{2}-\alpha\right)\hat{\psi}^{a,+}_{n+\alpha}-\frac{1}{2}\left(\hat{\psi}^{b,+}\phi^i\right)_{n+\alpha}(\sigma^i)^b_a
 \\
i\{R^i_n,\hat{\psi}^{\alpha}_{a,-}\}&=\frac{1}{2}\hat{\psi}^{n+\alpha}_{b,-}(\sigma^i)^b_a ~~~~~~~ 
i\{R^i_n,\hat{\psi}^{\alpha}_{a,+}\}=-\frac{1}{2}\hat{\psi}^{n+\alpha}_{b,+}(\sigma^i)^b_a \\
i\{R^i_n,R^j_m\}&=\frac{nk}{2}\delta_{n+m,0}+i \epsilon_{ijk}R^k_{n+m} \\
i\{L_n,R^j_m\}&=0\\
\{\hat{\psi}^{\alpha}_{a,+},\hat{\psi}^{\beta}_{b,-}\}&=\alpha^2k\delta_{\alpha+\beta}\delta_{ab}+L_{\alpha+\beta}\delta_{ab}+\frac{1}{2}(R^iR^i)_{\alpha+\beta}\delta_{ab}-(\alpha-\beta)R^i_{\alpha+\beta}(\bar{\sigma}^a_b)\;,  
\end{align*}
Where the modes are defined as follows:
\begin{align*}
L_n &= \int {\rm d}\theta e^{-in\theta} \mathcal{L_+}\;,\qquad\qquad
R^i_n = \int {\rm d}\theta e^{-in\theta} \phi^i \;,\\
\hat{\psi}^{a,+}_{\alpha} &= \int {\rm d}\theta e^{-i\alpha\theta} \psi^{a,-}\;,\qquad\quad 
\hat{\psi}^{a,-}_{\alpha} = \int {\rm d}\theta e^{-i\alpha\theta} \psi^{a,+}\;.\\
\end{align*}
By adding the Sugawara term
\begin{equation*}
L_n\rightarrow L_n'=L_n+\frac{1}{2}(R^iR^i)_n
\end{equation*}
the $i\{L_n,R^j_m\}$ gets modified as
\begin{equation*}
i\{\hat{L}_n,R^j_m\}=- mR^j_{n+m}\;,
\end{equation*}
and the supercharge anti-commutator takes the form:
\begin{equation*}
\{\hat{\psi}^{\alpha}_{a,+},\hat{\psi}^{\beta}_{b,-}\} =\alpha^2k\delta_{\alpha+\beta}\delta_{ab}+L^{'}_{\alpha+\beta}\delta_{ab}-(\alpha-\beta)R^i_{\alpha+\beta}(\bar{\sigma}^a_b) \;.
\end{equation*}
Note that the second and third term in the previous anti-commutator combined give the modified Sugawara generator $L'_{\alpha+\beta}$ so that the non-linear terms are absent in the final Poisson bracket.
Thus, we see that the asymptotic AdS algebra will not have any non-linearity in the R-symmetry charges. As a consequence, the corresponding asymptotic flat $\cN=8$ Super-BMS$_3$ algebra will also present no non-linearity.
\section{ AdS analysis and flat-space identifications}\label{AppC}

\begin{align*}\label{SCA}
[L_n,L_m]=&(n-m) L_{n+m}\;, \qquad
[L_n,R_p^{\alpha}]=\left( \frac{n}{2} - p  \right)R^{\alpha}_{n+p}\\
[L_n,T^a]=&0\;, \qquad \qquad\qquad\quad
\{R^{\alpha}_p,R^{\beta}_q\}= L_{p+q}\eta^{\alpha\beta}-\frac{\rm i}{6\hat{\alpha}}(p-q)(\lambda^a)^{\alpha\beta}T^a \delta_{p+q,0}\;, \\
[T^a,R_p^{\alpha}]=&{\rm i}(\lambda^a)^{\alpha}_{\beta}R_{p}^{\beta}\;, \qquad\qquad\;\;
[T^a,T^b]={\rm i}f^{abc}T^c\;,
\end{align*}
and similarly for the anti-chiral sector. The structure constants of the above algebra are the same as defined in section \ref{sec2}. We begin with two such identical copies of Superconformal algebras.
To get the asymptotic quantum algebra, let us begin with  the gauge fields and generic variation parameters for the two copies of AdS:
\begin{align*}
A&=\left[  L_1+\frac{r}{l}L_0+\left(\frac{r^2}{4l^2}-\frac{1}{2}\mathfrak{L}_+\right)L_{-1}+\mathfrak{A}Q_{\alpha}R^{-\alpha}+\frac{1}{2}\frac{k_l}{k_B}\phi^aT^a                                                               \right]{\rm d}x^++\frac{{\rm d}r}{2l}L_{-1}\;,
 \\
\bar{A}&=\left[  \bar{L}_{-1}-\frac{r}{l}\bar{L}_0+\left(\frac{r^2}{4l^2}-\frac{1}{2}\mathfrak{\bar{L}}_-\right)\bar{L}_{1}+\bar{\mathfrak{A}}\bar{Q}_{\alpha}\bar{R}^{+\alpha}+\frac{1}{2}\frac{k_l}{k_B}\bar{\phi}^a\bar{T}^a                                                               \right]{\rm d}x^-+\frac{{\rm d}r}{2l}\bar{L}_1\;,
\end{align*}
where $k_l=\frac{c}{6}$, where $c$ is the central charge of the quantum superconformal algebra.  \\
Asymptotic gauge transformations $\delta A=\delta \lambda+[A,\lambda]$ generate the asymptotic symmetries of the theory. The generic variation parameters are:
\begin{align*}
\lambda&=\chi^nL_n+\epsilon_{+,\alpha}R^{+,\alpha}+\epsilon_{-,\alpha}R^{-,\alpha}+\omega^aT^a\;, \\
\bar{\lambda}&=\bar{\chi}^n\bar{L}_n+\bar{\epsilon}_{+,\alpha}\bar{R}^{+,\alpha}+\bar{\epsilon}_{-,\alpha}\bar{R}^{-,\alpha}+\bar{\omega}^a\bar{T}^a\;. 
\end{align*}
\subsection*{AdS unbarred Sector Variation:}
Here we present the constraints on the parameters and the variations of the independent fields:
\begin{align*}
\chi^0=&\frac{r}{l}\chi_1-\chi_1' \;,
\\
\chi_{-1}=&-\frac{r}{2l}\chi_1'+\frac{1}{2}\chi_1''+\left(\frac{r^2}{4l^2}-\frac{1}{2}\mathfrak{L}_+\right)\chi_1+\frac{\mathfrak{A}}{2}Q_{\alpha}\epsilon^{\alpha}_+\;,
\\
\epsilon_{-,\alpha}=&-\epsilon_{+,\alpha}'+\mathfrak{A}Q_{\alpha}\chi_1+\frac{k_l}{2 k_B}\phi^a\epsilon_{+,\beta}(\lambda^a)^{\beta}_{\alpha}+\frac{r}{2l}\epsilon_{+,\alpha}\;,
\\
\delta\mathfrak{L}_+=&-\chi_1'''+\mathfrak{L}_+'\chi_1+2\mathfrak{L}_+\chi_1'-3\mathfrak{A}Q_{\alpha}\epsilon_{+,\alpha}'-\mathfrak{A}Q_{\alpha}'\epsilon_{+,\alpha} 
+\mathfrak{A}\frac{k_l}{ k_B}Q_{\alpha}\phi^a\epsilon_{+,\beta}(\lambda^a)^{\beta}_{\alpha}\;,
\\
\mathfrak{A}\delta Q_{\alpha}=&-\epsilon_{+,\alpha}'' +\mathfrak{A}Q_{\alpha}'\chi_1+\frac{3}{2}\mathfrak{A}Q_{\alpha}\chi_1'+\frac{k_l}{2 k_B}(\lambda^a)^{\beta}_{\alpha}\left[2\phi^a\epsilon_{+,\beta}'+(\phi^a)'\epsilon_{+,\beta}+\right]+\frac{1}{2}\mathfrak{L}_+\epsilon_{+,\alpha} 
\\ 
&-\mathfrak{A}\frac{k_l}{2 k_B}(\lambda_a)^{\beta}_{\alpha}\phi^aQ_{\beta}\chi_1-\frac{k_l^2}{4k_B^2}\phi^a\phi^b(\lambda^a)^{\gamma}_{\beta}(\lambda^b)^{\beta}_{\alpha}\epsilon_{+,\gamma}+\mathfrak{A} \omega^aQ_{\beta}(\lambda^a)^{\beta}_{\alpha}\;,
\\
\delta\phi^a=&2\frac{k_B}{k_l}(\omega^a )'+\phi^b\omega^c f^{abc}+2\mathfrak{A}Q_{\alpha}\epsilon_{+,\beta}(\lambda^a)^{\alpha\beta} \;.
\end{align*}
\subsection*{AdS Barred Sector Variation:}

Similar computations for the barred sector will give: 
\begin{align*}
\bar{\chi}_0=&-\frac{r}{l}\chi_{-1} +\bar\chi_{-1}'\;,
\\
\bar{\chi}_1=&-\frac{r}{2l}\bar{\chi}_{-1}'+\frac{1}{2}\bar{\chi}_{-1}''+\left(\frac{r^2}{4l^2}-\frac{1}{2}\bar{\mathfrak{L}}_{-1}\right)\bar{\chi}_{-1}-\frac{\bar{\mathfrak{A}}}{2}\bar{Q}_{\alpha}\bar{\epsilon}_{-}^{\alpha}\;,
\\
\bar{\epsilon}_{+,\alpha}=&\bar{\epsilon}_{-\alpha}'+\bar{\mathfrak{A}}\bar{\chi}_{-1}\bar{Q}_{\alpha}-\frac{r}{2l}\bar{\epsilon}_{-,\alpha}+\frac{1}{12\alpha}(\lambda^a)^{\beta}_{\alpha}\bar{\phi}^a\bar{\epsilon}_{-,\beta} \;,
\\
\delta\bar{\mathcal{L}}_-=&-\bar{\chi}_{-1}'''+\bar{\mathcal{L}}_-'\bar{\chi}_{-1}+2\bar{\mathfrak{L}}_{-1}\bar{\chi}_{-1}'+3\bar{\mathfrak{A}}\bar{Q}_{\alpha}(\bar{\epsilon}^{-,\alpha})'+\bar{\mathfrak{A}}\bar{Q}_{\alpha}'\bar{\epsilon}^{-,\alpha}+\bar{\mathfrak{A}}\frac{k_l}{k_B}(\lambda^a)^{\beta\alpha}\bar{Q}_{\alpha}\bar{\phi}^a\bar{\epsilon}_{-,\beta} \;,
\\
\bar{\mathfrak{A}}\delta \bar{Q}_{\alpha}=&\bar{\epsilon}_{-,\alpha}''+\bar{\mathfrak{A}}\bar{\chi}_{-1}\bar{Q}_{\alpha}'+\frac{3\bar{\mathfrak{A}}}{2}(\bar{\chi}_{-1})'\bar{Q}_{\alpha}+\frac{k_l}{2k_B}
(\lambda^a)^{\beta}_{\alpha}[2\bar{\phi}^a\bar{\epsilon}_{-,\beta}'+(\bar{\phi}^a)'\bar{\epsilon}_{-,\beta}] -\frac{1}{2}\bar{\mathfrak{L}}_-\bar{\epsilon}_{-,\alpha}
\\
 &+\bar{\mathfrak{A}}\frac{k_l}{2k_B}\bar{\phi}^a(\lambda^a)^{\beta}_{\alpha}\bar{\chi}_{-1}\bar{Q}_{\beta} 
+\frac{k_l^2}{4k_B^2}\bar{\phi}^a\bar{\phi}^b(\lambda^a)^{\beta}_{\alpha}(\lambda^b)^{\gamma}_{\beta}\epsilon_{-,\gamma}+\bar{\mathfrak{A}}\bar{\omega}^a(\lambda^a)^{\beta}_{\alpha}\bar{Q}_{\beta}\;,
\\
\delta\bar{\phi}^a=&2\frac{k_B}{k_l}(\bar{\omega}^a)'+\bar{\phi}^b\bar{\omega}^cf^{abc}-2\bar{\mathfrak{A}}\bar{Q}_{\alpha}\bar{\epsilon}_{-,\beta}(\lambda^a)^{\alpha\beta}\;.
\end{align*}
\subsection*{Identification with flat fields and generators:}
Using these above relations, one can find the corresponding constraints and variations for the gauge field $\mathcal{A}$ \eqref{GF} and gauge transformation parameter $\Lambda$ \eqref{Gpara} that gives the asymptotic symmetry for the $3$D flat space time. Specifically:
$$\mathcal{A}=A+\bar{A}\;, \qquad \Lambda= \lambda + \bar \lambda\;,$$
\begin{align*}
J_n &= L_n - \bar{L}_{-n}\;,\qquad
M_n = \frac{L_n + \bar{L}_{-n}}{l}\;,\qquad
r_{\pm,\alpha}^1 = \sqrt{\frac{2}{l}} R_{\pm,\alpha}\;, 
\\
r_{\pm,\alpha}^2 &= \sqrt{\frac{2}{l}} \bar{R}_{\pm,-\alpha}\;,\qquad
\mathcal{R}^a = T^a - \bar{T}^a\;,\qquad\quad\;\;
\mathcal{S}^a = \frac{T^a + \bar{T}^a}{l}\;.
\end{align*}
Using this identification the map for the charges is the following:
\begin{align*}
\mathcal{M} &= \mathcal{L}_+ +\mathcal{\bar{L}}_-\;, \qquad
\mathcal{N} = l(\mathcal{L}_+ -\mathcal{\bar{L}}_-)\;, \qquad
\psi_{\pm\alpha}^1 = \sqrt{\frac{l}{2}} Q_{\pm\alpha}\;,
\\ 
\psi_{\pm\alpha}^2 &= \sqrt{\frac{l}{2}} \bar{Q}_{\mp\alpha}\;, \qquad \;
\rho^a = \phi^a + \bar{\phi}^a\;, \qquad\;\;\qquad
\tilde{\phi}^a = l(\phi^a - \bar{\phi}^a)\;,
\end{align*}
and the parameters are scaled as:
\begin{align*}
\xi^n &= \frac{l}{2}(\chi^n + \bar{\chi}^{-n})\;,\qquad
\Upsilon^n = \frac{1}{2} (\chi^n -\bar{\chi}^{-n})\;,\qquad
\lambda_S^a = \frac{l}{2} (\omega^a + \bar{\omega}^a)\;,
\\
\lambda_R^a &= \frac{1}{2} (\omega^a - \bar{\omega}^a)\;,\qquad\;
\zeta_{\pm,\alpha}^1 = \sqrt{\frac{l}{2}} \epsilon_{\pm,\alpha}\;\qquad\qquad
\zeta_{\pm,\alpha}^2 = \sqrt{\frac{l}{2}} \bar{\epsilon}_{\pm,-\alpha}\;.
\end{align*}
The modes of the charges are defined as follows:
\begin{align*}
\mathfrak{J}_m &= \lim_{l \rightarrow \infty}(\mathcal{L}^+_m - \bar{\mathcal{L}}^-_m)\;,\qquad \;\;   \mathfrak{M}_n =\lim_{l \rightarrow \infty}\frac{1}{l}(\mathcal{L}^+_n+\bar{\mathcal{ L}}^-_{-n})  
 \\
S^a_n &= \lim_{l \rightarrow \infty} \frac{1}{l} (\phi^a_n + \bar{\phi}^a_{-n})\;,\qquad
R^a_n =\lim_{l \rightarrow \infty} (\phi^a_n - \bar{\phi}^a_{-n})\;,
\\
\psi^{1,\alpha}_{\pm} &= \lim_{l \rightarrow \infty} \sqrt{\frac{2}{l}} Q^{\alpha}_{\pm}\;,\qquad\qquad \psi^{2,\alpha}_{\pm} = \lim_{l \rightarrow \infty} \sqrt{\frac{2}{l}} \bar{Q}^{\alpha}_{\mp}\;,
\\
c_J&=\lim_{l \rightarrow \infty}(c-\bar{c})\;,\qquad\qquad\quad c_M=\lim_{l \rightarrow \infty}\frac{1}{l} (c+\bar{c})\;.
\end{align*}
~~\\
Using these identifications, the final Asymptotic symmetry algebra for flat $3$D space time has been obtained in \eqref{UBMS}.

\section{Asymptotic gauge Field and gauge parameter for maximal extended super-BMS$_3$}\label{AppD}
The asymptotic gauge field and transformation parameter of \eqref{eq:auaphi} and \eqref{GGpara} need to be modified for the right asymptotic algebra \eqref{BMS}, as mentioned in sections \ref{sec2.3} and \ref{sec3}. The modified most generic gauge field are:
\begin{align}
a_u&=M_1-\frac{1}{4}\left( \mathcal{M}-\frac{1}{48\hat\alpha}\rho^a\rho^a       \right)M_{-1}+\frac{1}{24\hat\alpha}\rho^aS^a \;, \nonumber \\
a_{\phi}&=J_1-\frac{1}{4}\left( \mathcal{M}-\frac{1}{48\hat\alpha}\rho^a\rho^a       \right)J_{-1} - \frac{1}{4}\left(      \mathcal{N}-\frac{1}{24\hat\alpha}\tilde{\phi}^a \rho^a\right)M_{-1} \nonumber \\
&\quad+\mathfrak{A}\psi^1_{\alpha}r^{-,\alpha}_1-\bar{\mathfrak{A}}\psi^2_{\alpha}r_2^{-,\alpha} +\frac{1}{24\hat\alpha}\rho^a \mathcal{R}^a+\frac{1}{24\hat\alpha} \tilde{\phi}^a_{0} \mathcal{S}^a\;.  \nonumber 
\end{align}
 Here, we find the modified gauge transformation parameter that finally gives us \eqref{BMS}. The most generic transformation parameter has the form ,
\begin{equation}\label{p}
\Lambda=\xi^n_0M_n+\Upsilon^nJ_n+\tilde{\lambda}^a_{S0}\mathcal{S}^a + \tilde{\lambda}^a_R\mathcal{R}^a + \zeta^{1,\alpha}_{\pm} r^{1,\alpha}_{\pm} + \zeta^{2,\alpha}_{\pm} r^{2,\alpha}_{\pm}\;,
\end{equation}
The gauge field and gauge transformation parameter are constrained by equations of motion and gauge variation equations given as, $$da+ \frac{1}{2}[a,a]=0 , \quad \delta a= d \Lambda + [a, \Lambda].$$
The equation motion implies following relations : $$
\partial_\varphi\left( \mathcal{M}-\frac{1}{48\hat\alpha}\rho^a\rho^a  \right) =\partial_u\cN, \quad  \partial_u\mathcal{M}=0, \quad \partial_u\,\rho^a=0\;, \quad
\partial_u\psi^1_\pm=0\;,\quad \partial_u\psi^2_\pm=0\;, \quad \partial_\varphi \rho^a=\partial_u\tilde\phi^a_0.
$$
Here $\tilde\phi^a$ is the $u$ independent part of $\tilde\phi^a_0$. The gauge variation equation along $\phi$-direction provides following constrains and variation equations:
\begin{align}
\Upsilon^0&=-(\Upsilon^1)' \;, ~~~~~~~~~~~~~~~~~~~~\xi^0_0 =-(\xi^1_0)' \;,\nonumber
\\
\Upsilon^{-1}&=\frac{1}{2}\left[   (\Upsilon^1)''  -\frac{1}{2}\Upsilon^1 \left(  \mathcal{M}-\frac{1}{48\hat\alpha} \rho^a \rho^a      \right)                                                         \right]  \;,\nonumber
\\
\xi^{-1}_0&=-\frac{1}{2}  \left[  -(\xi^1_0)''+\frac{1}{2} \Upsilon^1\left(   \mathcal{N} - \frac{1}{24\hat\alpha}\rho^a  \tilde{\phi}^a              \right)     + \frac{1}{2} \xi^1_0 \left( \mathcal{M} - \frac{1}{48\hat\alpha}\rho^a\rho^a \right) -\mathfrak{A} \psi^1_{\alpha} \zeta^{1,\alpha}_{+} -\mathfrak{\bar{A}} \psi^2_{\alpha} \zeta^{2,\alpha}_{+}     \right] \;,\nonumber
\\
\zeta^{1,\alpha}_{-} &= -(\zeta^{1,\alpha}_{+})' +\mathfrak{A} \psi^{1}_{\alpha} \Upsilon^1 -\frac{1}{24\hat{\alpha}} i(\lambda^a)^{\beta}_{\alpha} \rho^a (\zeta^{1,\beta}_{+}) \\
\zeta^{2,\alpha}_{-} &= -(\zeta^{2,\alpha}_{+})' +\bar{\mathfrak{A}} \psi^{2}_{\alpha} \Upsilon^1 +\frac{1}{24\hat{\alpha}} i(\lambda^a)^{\beta}_{\alpha} \rho^a (\zeta^{2,\beta}_{+}) \nonumber\\
\delta \mathcal{M}&=   \left(\frac{1}{24\alpha}\rho^a \delta \rho^a \right)-4  (\Upsilon^{-1})'- \left(\mathcal{M}-\frac{1}{48\hat\alpha}\rho^a \rho^a \right) \Upsilon^0  \;,\nonumber
\\
\delta\mathcal{N}&=\frac{1}{24\hat\alpha}\delta(\tilde{\phi}^a \rho^a) -4(\xi^{-1}_0)'-\left( \mathcal{N}-\frac{1}{24\hat\alpha} \tilde{\phi}^a  \rho^a        \right)\Upsilon^0- \left(      \mathcal{M}-\frac{1}{48\hat\alpha}\rho^a  \rho^a      \right) \xi^0_0 \nonumber\\
&+4 \mathfrak{A} \psi^1_{\alpha} \zeta^{1,\alpha}_{-} +4 \mathfrak{\bar{A}} \psi^2_{\alpha} \zeta^{2,\alpha}_{-}   \;,\nonumber
\\
\delta \rho^a&=24\hat\alpha\tilde{(\lambda^a_R)}'+i \rho^b\tilde{\lambda^c_R}f^{abc} \;,\nonumber
\\
\frac{1}{24\hat{\alpha}}\delta \tilde{\phi}^a_0 &= \tilde{(\lambda^a_{S0})}'+i\frac{1}{24\hat{\alpha}}\left(\rho^b\tilde{\lambda^c_{S0}}f^{abc}-\tilde{\lambda^b_R}\tilde{\phi}^c_0 f^{abc} \right)\;,\nonumber
\\
\mathfrak{A} \delta \psi^1_{\alpha} &= (\zeta^{1,\alpha}_{-})' + \frac{1}{4} \left(   \mathcal{M} - \frac{1}{48\hat{\alpha}} \rho^a \rho^a  \right) \zeta^{1,\alpha}_{+} + \mathfrak{A} \psi^1_{\alpha} \Upsilon^0 + \frac{1}{24\hat{\alpha}} \rho^a \zeta^{1,\beta}_{-} i(\lambda^a)^{\beta}_{\alpha} -i \mathfrak{A} (\lambda^a)^{\beta}_{\alpha} \psi^1_{\beta} \tilde{\lambda}^a_R \nonumber
\\
\bar{\mathfrak{A}} \delta \psi^2_{\alpha} &= (\zeta^{2,\alpha}_{-})' + \frac{1}{4} \left(   \mathcal{M} - \frac{1}{48\hat{\alpha}} \rho^a \rho^a  \right) \zeta^{2,\alpha}_{+} + \bar{\mathfrak{A}} \psi^2_{\alpha} \Upsilon^0 - \frac{1}{24\hat{\alpha}} \rho^a \zeta^{2,\beta}_{-} i(\lambda^a)^{\beta}_{\alpha} +i \bar{\mathfrak{A}} (\lambda^a)^{\beta}_{\alpha} \psi^2_{\beta} \tilde{\lambda}^a_R \nonumber
\end{align}
Similarly, the gauge variation equation along $u$ gives us
 $$\partial_u\xi^1_0=\partial_\varphi\U^1,\quad \partial_u\U^n=0,\quad \partial_u\z^{1,\a}_\pm=0,
\quad \partial_u\z^{2,\a}_\pm=0, \quad \partial_u\tilde\l^a_{S0}=\partial_\varphi\tilde\l^a_R\;,\quad \partial_u\,\tilde\l^a_R=0.$$

Thus we see that other than $\xi^1_0,\Upsilon^1,\z^{1,\a}_+,\z^{2,\a}_+$ fields, the remaining components of $\Lambda$ are very much gauge field dependent. Further the variation of the charge reads
\begin{align}
\delta C=&-\frac{k}{4\pi}\int d\phi \langle\Lambda,\delta a_{\phi}\rangle \nonumber\\
=&-\frac{k}{4\pi}\int d\phi \left[ \frac{1}{2}\xi^1\delta\mathcal{M}+\frac{1}{2}\Upsilon^1\delta \mathfrak{J} +   \frac{1}{2}(\delta \tilde{\phi}^a\tilde\lambda^a_R+\delta \rho^a\tilde\lambda^a_S)  + \mathfrak{A} \delta\psi^1_{\alpha} \zeta^{1,\alpha}_{+}   + \mathfrak{\bar{A}} \delta\psi^2_{\alpha} \zeta^{2,\alpha}_{+}          \right] \nonumber\\
&+\frac{k}{4\pi}\int d\phi\frac{1}{48\alpha} \left[   \rho^a\delta \rho^a\xi^1+\tilde{\phi}^a\delta \rho^a \Upsilon^1  +\rho^a\delta \tilde{\phi}^a \Upsilon^1                         \right],
\end{align}
%\nonumber\\
%&-\frac{k}{4\pi}\int d\phi\frac{1}{2} \left[C\tilde{\phi}^a\delta \tilde{\phi}^a+D \rho^a\delta \tilde{\phi}^a + A\tilde{\phi}^a\delta \rho^a  + B \rho^a \delta \rho^a                               \right]
%\end{align}
where $\xi^1,\mathfrak{J},\tilde\lambda^a_S$ are the $u$-independent parts of $\xi^1_0,\mathcal{N},\tilde\lambda^a_{S0}$ respectively.
The above expression can be integrated to compute the charge by defining 
\begin{align}
{\lambda}^a_S&=\tilde\lambda^a_S-A\tilde{\phi}^a-B\rho^a , \quad
{\lambda}^a_R=\tilde\lambda^a_R-C\tilde{\phi}^a-D\rho^a 
\end{align}
to be gauge independent parameters where,
\begin{align}
A=D=\frac{1}{24\hat\alpha}\Upsilon^1, ~~~~~~~~~B=\frac{1}{24\hat\alpha}\xi^1,  ~~~~~~~~~ C=0,
\end{align}
and considering all the independent components $\xi^1,\Upsilon^1,{\lambda}^a_R,{\lambda}^a_S,\zeta^{1,\alpha}_{+},\zeta^{1,\alpha}_{-}$ not to vary at the boundary. Thus the final variation of the charge looks like 
\begin{align}
\delta C=-\frac{k}{4\pi}\int d\phi \left[ \frac{1}{2}\xi^1\delta\mathcal{M}+\frac{1}{2}\Upsilon^1\delta \mathfrak{J} +   \frac{1}{2}(\delta \tilde{\phi}^a\lambda^a_S+\delta \rho^a\lambda^a_R) + \mathfrak{A} \delta\psi^1_{\alpha} \zeta^{1,\alpha}_{+}   + \mathfrak{\bar{A}} \delta\psi^2_{\alpha} \zeta^{2,\alpha}_{+}       \right] .
\end{align}
The above expression can be integrated out to get the charges and it
can be checked that the above charge rightly reproduces the algebra \eqref{BMS}.  
Finally, inserting back all the constraints, we get the expression for the transformation parameter:
\begin{align}\label{FP}
\Lambda=&\xi^1M_1+\Upsilon^1J_1+\left(    \lambda^a_R+\frac{1}{24\hat\alpha} \Upsilon^1 \tilde{\phi}^a                        \right)\mathcal{R}^a   
+ \left(  \lambda^a_S+\frac{1}{24\hat\alpha}\Upsilon^1 \tilde{\phi}^a       +\frac{1}{24\hat\alpha}\xi^1  \rho^a \right) \mathcal{S}^a \nonumber\\
& -(\xi^1)'M_0-(\Upsilon^1)'J_0 
+\frac{1}{4}\left[  2(\Upsilon^1)''  -\Upsilon^1 \left(   \mathcal{M}-\frac{1}{48\hat\alpha} \rho^a \rho^a                      \right)  \right]J_{-1}  \nonumber\\
&-\frac{1}{4} \left[    -2(\xi^1)''+  \Upsilon^1\left( \mathcal{N} -\frac{1}{24\hat\alpha}   \tilde{\phi}^a \rho^a   \right)    +\xi^1    \left(   \mathcal{M}-\frac{1}{48\hat\alpha} \rho^a  \rho^a                                   \right)                       \right]      M_{-1}  \nonumber \\
 & + \zeta^{1,\alpha}_{+}  r^{1,\alpha}_{+} +  \zeta^{2,\alpha}_{+}  r^{2,\alpha}_{+} 
+ \left(   -(\zeta^{1,\alpha}_{+})' + \mathfrak{A} \psi^1_{\alpha} \Upsilon^1 - \frac{1}{24 \hat{\alpha}} i (\lambda^a)^{\beta}_{\alpha} \rho^a (\zeta^{1,\beta}_{+})           \right) r^{1,\alpha}_{-} \nonumber\\
&+\left(   -(\zeta^{2,\alpha}_{+})' + \bar{\mathfrak{A}} \psi^2_{\alpha} \Upsilon^1 + \frac{1}{24 \hat{\alpha}} i (\lambda^a)^{\beta}_{\alpha} \rho^a (\zeta^{2,\beta}_{+})           \right) r^{2,\alpha}_{-}.             
\end{align}

\bibliography{bms3}
\bibliographystyle{jhep}

\end{document}